\documentclass{JINST}
\usepackage{epsfig}
\usepackage{graphicx}
\usepackage{array}
\usepackage{float}
\usepackage{rotfloat}
\usepackage{amstext}
\usepackage{graphicx}
\usepackage{rotating}
\usepackage{amssymb}
\usepackage[utf8]{inputenc}

\title{Micromegas-TPC operation at high pressure in xenon-trimethylamine mixtures }

\author{S. Cebrián$^a$, T. Dafni$^a$, E. Ferrer-Ribas$^b$, I. Giomataris$^b$, D. Gonzalez-Diaz$^{a,c}$,
 H.~Gómez $^{a,d}$, D.C. Herrera$^{a,}$\footnote{Corresponding autor} , F.J. Iguaz$^a$, I.G. Irastorza$^a$, G. Luzon$^a$, A. Rodríguez$^a$,
 L.~Segui$^a$, A. Tomás$^a$ \\
\llap{$^a$}Laboratorio de Física Nuclear y Astropartículas, Universidad de Zaragoza\\
  C/ Pedro Cerbuna 12, 50009, Zaragoza, Spain\\
\llap{$^b$}IRFU,Centre d’Etudes Nucleaires de Saclay (CEA-Saclay)\\
  91191 Gif sur Yvette, France\\
\llap{$^c$}Department of Engineering Physics, Tsinghua University, Liuqing building, 10084, Beijing, China\\
\llap{$^d$}Laboratoire de l'Accélérateur Linéaire (LAL), Université Paris-Sud 11, Bâtiment 200
91898 Orsay Cedex (FRANCE)\\
 E-mail: \email{diana.he@unizar.es}}

\abstract{In this work we present a systematic study of Micromegas detectors
in high pressure gaseous Xenon using trimethylamine
(TMA) as quencher gas. Gas gains and energy resolutions for $22.1$ keV X-rays
are measured for pressures between 1 and
10 bar and various relative concentrations of TMA from 0.3 $\%$ to 15 $\%$. We
observe stable operation at all pressures, and a strongly enhanced gas gain,
suggestive of Penning-like energy-transfer processes. 
The effect is present at all pressures and it is strongest at TMA concentrations
ranging from 1.5 $\%$ to 3 $\%$. Operating in this concentration range, the maximum gain 
reached values as high as $\sim2\times10^{3}$ ($\sim5\times10^{2}$) at 1 (10)
bar. Besides, the energy resolution achievable for $22.1$ keV X-rays is substantially
better than the one previously obtained in pure Xe, going down to
7.3 $\%$ (9.6 $\%$) FWHM for 1 (10) bar. 
These results are of interest for calorimetric applications of high pressure gas Xe TPCs,
in particular for the search of the neutrinoless double beta decay of
$^{136}$Xe. The resolutions achieved would extrapolate into 0.7 $\%$
(0.9 $\%$) FWHM at the $Q_{\beta\beta}$ value of $^{136}$Xe for
1 (10) bar.}


\begin{document}
\maketitle

\section{Introduction}

Xenon-filled gaseous Time Projection Chambers (TPCs) are considered
an interesting alternative to liquid Xe detectors in applications
for which enhanced topological information, as well as energy resolution,
are required. They have found successful applications, e.g., in X-rays
astronomy and in double beta  decay searches ($\beta\beta$). The approach
of using gas TPCs in $\beta\beta$ searches was pioneered by the Gothard
TPC in the 1990's \cite{Gothard}. Much more
recently, the NEXT experiment is following a similar approach and 
is constructing a 100 kg gas Xe TPC for this goal \cite{NEXT}.
The NEXT readout concept relies on the detection of the scintillation
of pure Xe for both $t_{0}$ determination and enhanced energy resolution.
Another collaboration, EXO-gas, also develops a gas TPC to search
for the double beta decay of $^{136}$Xe, and considers several options
for its readout \cite{EXO}.

More generically, the technical complexity of gas TPCs has been partially
alleviated by the advent of micropattern gas detectors (MPGD), simpler  
and with better scaling-up prospects than conventional multiwire proportional
chambers (MWPC). One of the most attractive MPGD from the point of
view of energy resolution are the Micromesh Gas Structures (Micromegas) \cite{Giomataris}.
These readouts are object of active development for application to
rare event searches within the context of the R\&D project T-REX \cite{T-REX}.
In particular, Micromegas fabricated with the $\textit{microbulk}$
technique \cite{Microbulk} have shown
particularly good prospects regarding aspects like radiopurity \cite{Radiopurity},
or energy resolution \cite{EnReMi}, both
aspects crucial for application to $\beta\beta$ searches. The results
presented in this paper are a further step in the studies of
energy resolution with this type of readouts.

Micromegas readouts make use of a metallic micromesh suspended over
an anode plane by means of insulator pillars, defining an amplification
gap of the order of 50-150 $\mu$m. It is known \cite{Giomataris99}
that the way the amplification develops in a Micromegas gap is such
that its gain $M$ is less dependent on geometrical factors (the gap
size) or environmental ones (like the temperature or pressure of the
gas) than other MPGDs or MWPCs. In addition, the amplification in
the Micromegas gap has less inherent statistical fluctuations than
that of wires, due to the faster transition from the drift field to
the amplification field provided by the micromesh \cite{MicroFaster}.
Micromegas of the $\textit{microbulk}$ type are built out of a double-clad
kapton foil, by applying an appropriate chemical bath that removes both copper and
kapton to form the Micromegas holes \cite{MicroTech}. The homogeneity of the gap in
these structures is superior to other more conventional Micromegas,
further improving gain stability and homogeneity and therefore energy
resolution. Experimentally, Full Width at Half Maximum (FWHM) resolutions of $\sim$11$\%$ for
the 5.9 keV peak of $^{55}$Fe in Ar-based mixtures are now routinely
achieved by the latest generation of these detectors.

The work performed up to now has been focused on establishing the
capability of microbulk detectors to work in high pressure pure Xe,
and more specifically to measure their energy resolution in those
conditions. In our previous works \cite{Paco}
we have experimentally demonstrated that operation of $\textit{microbulk}$
Micromegas in pure noble gases is feasible, even at high pressures,
and reaching gains above 100 for the case of Xe. This is a remarkable
result because in absence of quencher early breakdown occurs at high pressure
in all other MPGDs \cite{Coimbra}. We tentatively
attribute this result to the ``geometrical quenching'' achieved by
the fact that the avalanche in $\textit{microbulk}$ detectors happens
totally enclosed in a kapton cell, preventing to some extent the propagation
of photon-driven avalanches.
Geometrical streamer-quenching is indeed a documented phenomenon, 
and has been used with limited success in some non-standard implementations of Resistive Plate Chambers \cite{Diego}.

In this work we present results obtained
with trimethylamine (TMA) as a quencher for Xe. The mixture Xe + TMA
was studied in the past for wire chambers 
at 1 bar \cite{XeBas}. The authors of \cite{XeBas} have shown that the addition 
of TMA to Xe at a given field,
substantially increases the detector gain with respect to the
one in pure Xe, a phenomenon attributed to Penning effect. 
On the other hand, improvements of $\sim25\,\%$ in energy resolution have also also 
observed compared with the conventional non-Penning mixtures. 
We extend here that study to higher pressures up to 10~bar using Micromegas detectors. 
We perform systematic measurements of gas gain and energy resolution  
for different TMA fractions, from $0.3$ $\%$ to 15$\%$.
In general, the operation of the detector in Xe+TMA mixtures
improves substantially the gas gain as well as the energy resolution, 
up to concentrations of the quencher in the range $\sim 2-3$ $\%$ 
and for all pressures tested. The improvement is specially outstanding at high
pressures, given that operation in pure Xe degrades rapidly with pressure \cite{Coimbra}.
We believe that this systematic study could provide a starting point for evaluating the recent proposal of
using TMA as a means to reduce the Fano factor in Xe-based chambers \cite{Nygren}.

The article is structured as follows: in section 2 we describe the
experimental setup and in section 3 the experimental procedure. In section
4 we present the results obtained and discuss them. We finish in section
5 with our conclusions.

\section{Experimental setup}

\begin{figure}[hb]
\begin{centering}
\includegraphics[scale=0.35]{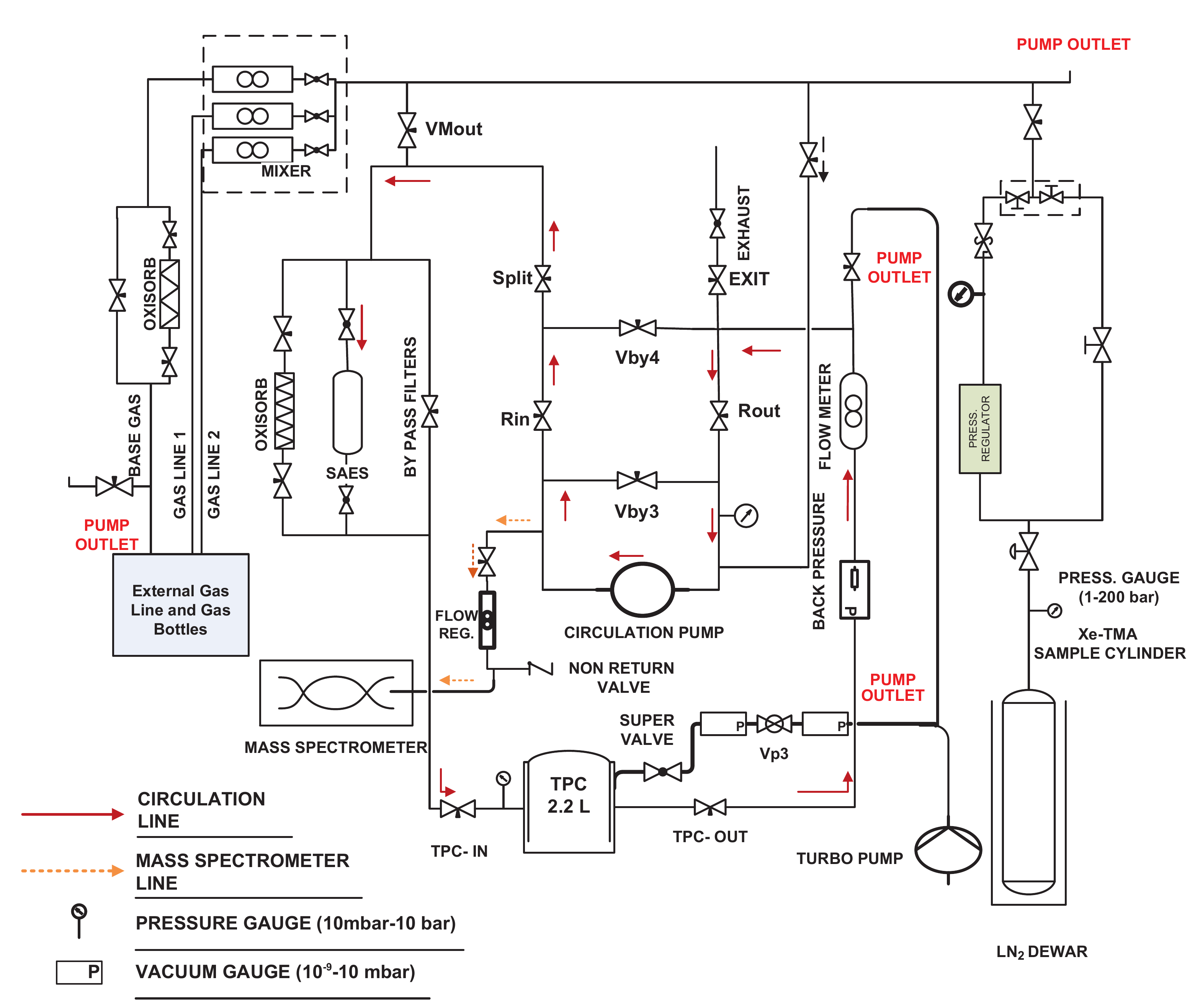}
\par\end{centering}
\caption{Schematic view of the experimental setup with the location of the TPC,
the Xe-TMA sample cylinder and the vacuum, filling, purification, recovery
and mass spectrometer lines. \label{fig:ExperimentalSetup}}
\end{figure}

The experimental setup used in the present investigation is similar
to the one used in our earlier studies in pure Xe and has been described
in detail in a previous paper \cite{Paco}. It consists
of a cylindrical stainless steel chamber and a gas system. The gas system includes: a filling gas line, a recirculation gas line, a gas recovery line and a mass spectrometer line. 
The last two parts have been added to the previous setup. A schematic view is shown in figure \ref{fig:ExperimentalSetup}.
 
The chamber has an active volume of $2.2$ liters (10 cm  in height, 16 cm in diameter) constructed 
with ultra-high vacuum specifications and is rated for operation at 15 bar. Inside the chamber,
two circular copper plates  ($10$ cm in diameter) form the drift/conversion region of $1$ cm. 
The cathode (top plate) is connected to negative voltage.
Centered relative to the bottom plate, a Micromegas is located, having $35$\,mm diameter, $50\,\mu$m 
gap, with $50$\,$\mu$m diameter holes and $115\,\mu$m pitch. 
The detector was built with the microbulk technique \cite{Microbulk}.
The mesh is grounded to the body chamber while a positive voltage is applied to the anode. 
The signal from the anode is fed into a CANBERRA preamplifier+amplifier chain (model 2005 and 2022 respectively), 
the latter with a shaping time of $4$ $\mu$s. The amplified signal is 
finally recorded by a multichannel analyzer (AMPTEK MCA 8000A), that produces the pulse height distribution.    

The gas filling line and recirculation line, including a membrane pump (circulating process), several valves, gauge pressures, and filters
are sketched in detail in figure \ref{fig:ExperimentalSetup}. In addition, 
a new purifier has been installed (SAES 702), which removes water vapor and electronegative impurities (H$_{2}$O, O$_{2}$, CO$_2$) and 
it is compatible with TMA.
The vacuum system includes a turbo-molecular pump, 
two gauge pressures and four outlets that permit to attain vacuum values down to 10$^{-6}$\,mbar.

The gas recovery system has a stainless steel sample cylinder of $2.2$ liters 
that is immersed in a Dewar flask filled with liquid nitrogen (LN$_2$) during the recovery process. 
The outlet of the sample cylinder is connected to two gas lines: one of them for the gas recovery and the other one for the filling of the TPC.
On the other hand, a Pfeiffer OmniStar mass spectrometer has been added to the gas system via a low pressure gas line. 
The mass spectrometer is used to quantify the gas composition of
Xe+TMA mixtures and to monitor the electronegative impurities. 
Measurements filling the gas system at $2$ bar with a pre-mixed $75\,\%$ Xe+$25\,\%$ TMA mixture 
were performed in order to establish the calibration factor needed for determining the amount of TMA in each 
mixture used. In case of water vapor and oxygen only relative changes could be estimated.

\section{Experimental procedure and method of analysis}

All measurements were carried out with a $^{109}$Cd source collimated with a $2$ mm diameter spot, 
that was placed on the center of the cathode plate. The chamber and the gas system were
pumped to values down to $10^{-6}$\,mbar. 
The chamber was routinely baked out at a temperature of $100^{\circ}$C 
during three hours, obtaining outgassing rates below $5\times10^{-5}$\,mbarl/s, before the gas filling.
The Xe used had a purity-grade of 6. At the beginning of all measurements,
a Xe+TMA mixture with high concentration ($\sim 7$\,\%) of TMA was prepared in the sample cylinder.
Lower concentrations of the additive gas were obtained adding fresh Xe from the external bottle. 
After the preparation of every new mixture, we observed that the circulation of the mixture through the filter produced changes in the 
TMA concentration, that is due to the SAES filter absorbing or expelling TMA depending on the previous gas mixture used. 
This fact forced us to monitor the TMA concentration obtained after a certain stabilization time, 
that was never exceeding 1h. The procedure was as follows: the chamber was filled at the desired pressure, 
and the circulation process through the purifier started, keeping the pressure constant. 
Secondly, energy spectra were acquired every two minutes at a gas gain above $100$ and a reduced drift field below $100$ V/cm/bar.
Then the gas gain and energy resolution were monitored until the variations seen were less than $5\,\%$.
Such stationary conditions were typically reached within some $20$ minutes, 
a situation that we interpret as the mixture being homogeneous throughout the complete gas system. 
This fact was verified several times, 
monitoring the TMA concentration with the mass spectrometer at the same time.

\begin{figure}[h]
\begin{centering}
\includegraphics[scale=0.45]{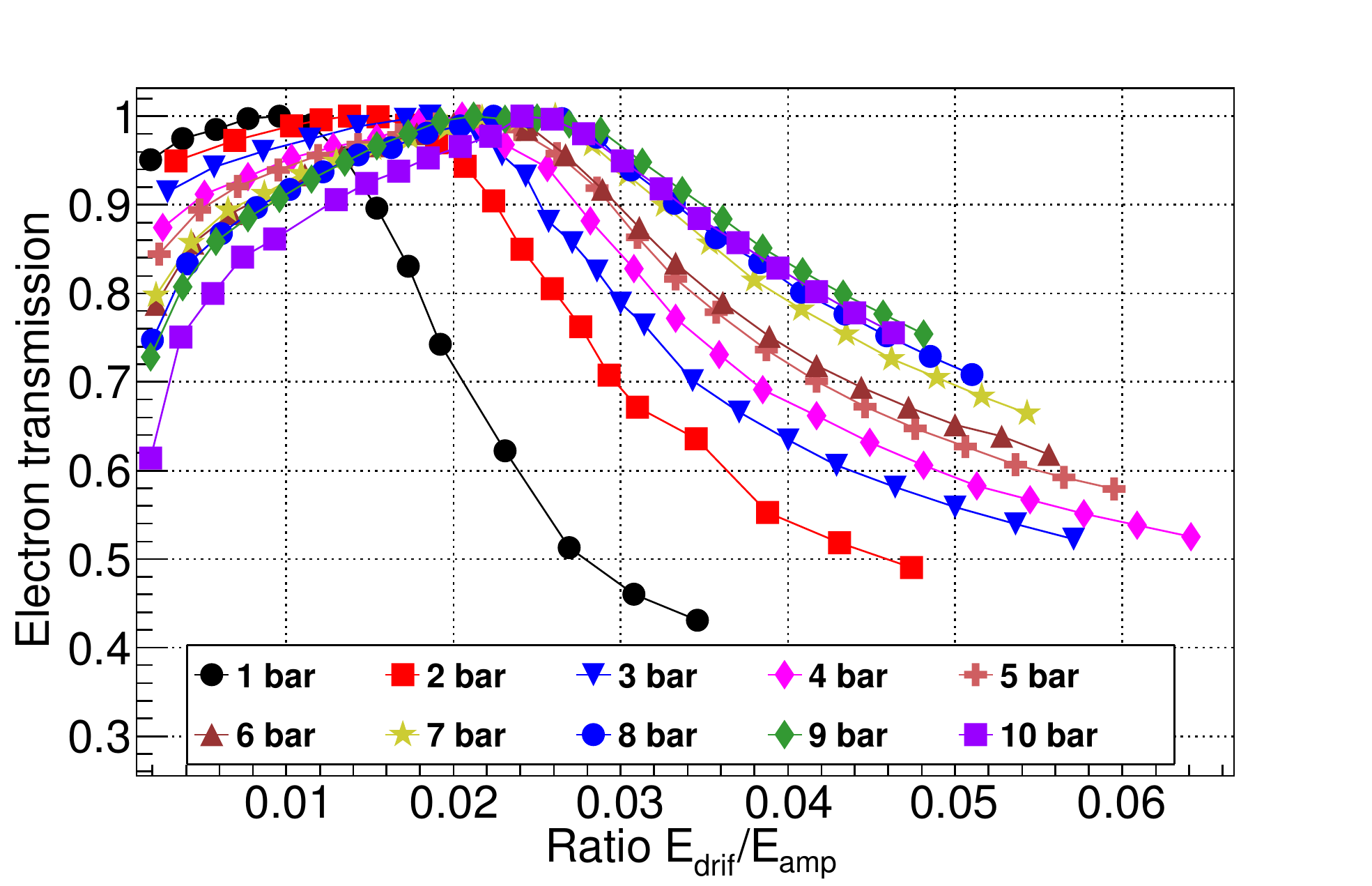}
\par\end{centering}
\caption{
Electron transmission of the Micromegas detector as a function of the ratio of drift-to-amplification fields in Xe+TMA mixtures, for pressures between $1$ and $10$ bar.
The percentage of TMA for each pressure is around 
$1.5$\,\%. The operation point was selected within the plateau region.\label{fig:ElectronTransmission}}
\end{figure}

Once the gas mixture was stabilized, the operating point was established by ensuring the 
maximum electron transmission of the Micromegas mesh.
For this goal the amplification voltage was kept constant for a gas gain around $300$
while the drift voltage was scanned within a range  
$0.1$-$1$ kV/cm ($1$ bar) and $0.2$-$5$ kV/cm ($10$ bar).
Figure \ref{fig:ElectronTransmission} shows the typical curves of  
electron transmission as a function of the ratio between drift and
amplification fields for pressures from $1$ to $10$ bar using mixtures with around $1.5\,\%$ TMA,
where the values have been normalized to the maximum of each curve. 
Since the electron transmission depends on
the ratio of drift-to-amplification fields $E_{drift}/E_{amp}$, as it is showed in figure 
\ref{fig:ElectronTransmission} and in earlier works \cite{Giomataris, Paco},
it was kept constant during measurements.

The increase in the amplification field was stopped when consecutive sparks occurred in a short time ($30$ s),  
typically the variations were between $34$ ($84$) and $58$ ($114$) kV/cm for $1$ ($10$) bar.
In particular, at high pressures where it is necessary to apply high fields,
the increase was until a spark occurred in order to avoid irreversible damage. 
At the end of each set of measurements, the TMA concentration was measured with the mass spectrometer.
Finally, the chamber and the gas system was cleaned by recovering the gas and then pumping it. 
The recovery process was performed by cooling the sample cylinder to LN$_2$ temperature (\ref{fig:ExperimentalSetup}).


%

\begin{figure}[hb]
\begin{centering}
\includegraphics[scale=0.40]{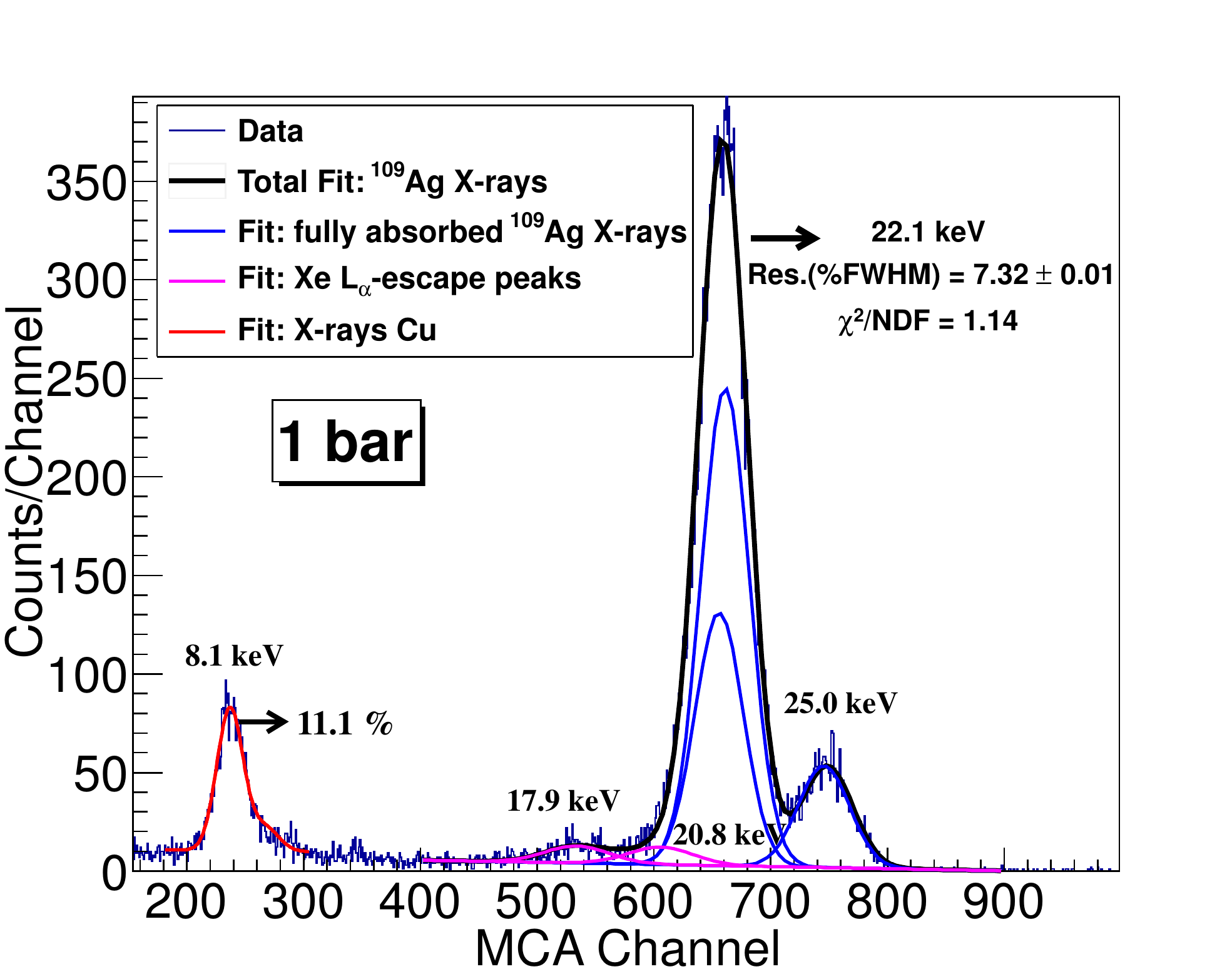}\includegraphics[scale=0.40]{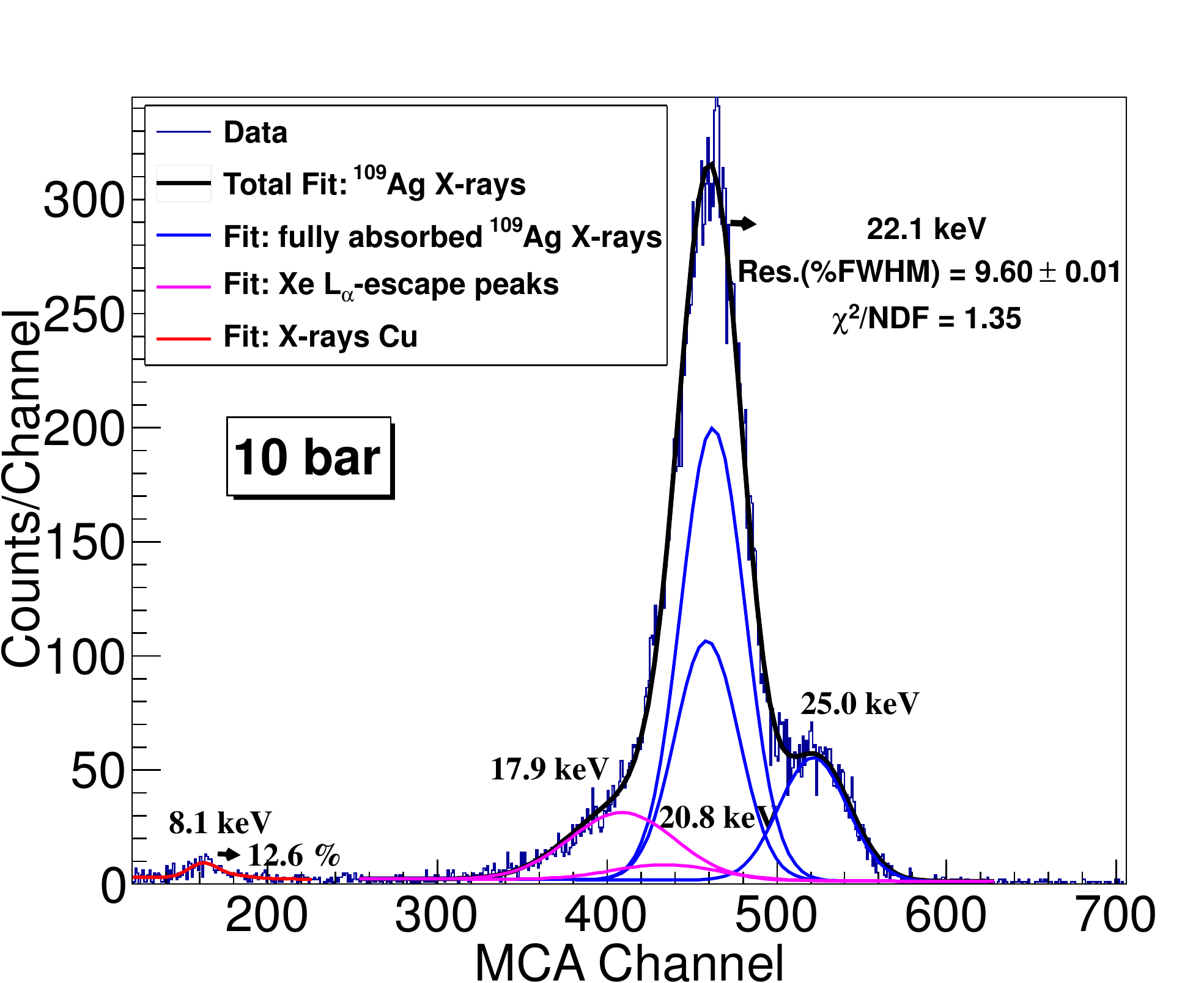}
\par\end{centering}
\caption{Characteristic X-rays energy spectra originated from $^{109}$Cd source, in a Xe+$1.7$ \% TMA mixture at 1 bar (left) and $1.1$\% TMA at
$10$ bar (right). The fit performed to the overall $^{109}$Ag K-fluorescence lines consists in a 3-step routine that is shown in each spectrum. In
addition, the Cu K-fluorescence at $8.1$ keV is separately fitted to a single Gaussian.
\label{fig:Energy-X-rays-spectra-1-and-10bar}}
\end{figure}

In the offline analysis, the peak at $22.1\,$keV was used to obtain the
gas gain and energy resolution. As the mean energy for the formation of an 
ion pair (W) has not been measured for Xe+TMA mixtures, 
the value of $22$ eV for pure Xe was used for all gas gains calculations; 
the estimated error due to this assumption is below $10\,\%$.
In figure \ref{fig:Energy-X-rays-spectra-1-and-10bar}, the typical energy spectrum acquired with 
a $^{109}$Cd source is shown at 1 bar in a Xe+$1.7\%$\,TMA mixture 
(left) and at 10 bar with $1.1$\%\,TMA (right). 
In both spectra the K$_{\alpha}$ and K$_{\beta}$ lines from Ag fluorescence are clearly distinguished.
The corresponding escape peaks from Xe are observed below the K$_{\alpha}$ line, 
located at $17.9$ keV and $20.8$ keV.
The Cu K-fluorescence at $8.1$ keV is also observed in the spectra.
These events are produced from the interaction of X-rays with the electrodes of the Micromegas. 

The Ag K$_{\alpha}$, K$_{\beta}$ lines and the corresponding Xe escape peaks (L-shell) were 
fitted in the energy range between $14$ and $30$\,keV using a $3$-step routine. 
The results of this procedure are shown in figure \ref{fig:Energy-X-rays-spectra-1-and-10bar}.  
In the first step, the complete range was fitted to one Gaussian function K$_{\alpha1}$($22.2$\,keV) over a linear background. In 
the second one, two Gaussian functions were added corresponding to K$_{\alpha2}$($22.0$ keV) and K$_{\beta}$($25.0$\,keV) lines.
In the last step, two more Gaussian functions were added which correspond to the Xe X-rays escape peaks. 
The input parameters in the last two steps are the calculated ones at previous steps. 

%
\begin{figure}[hb]
\begin{centering}
\includegraphics[scale=0.7]{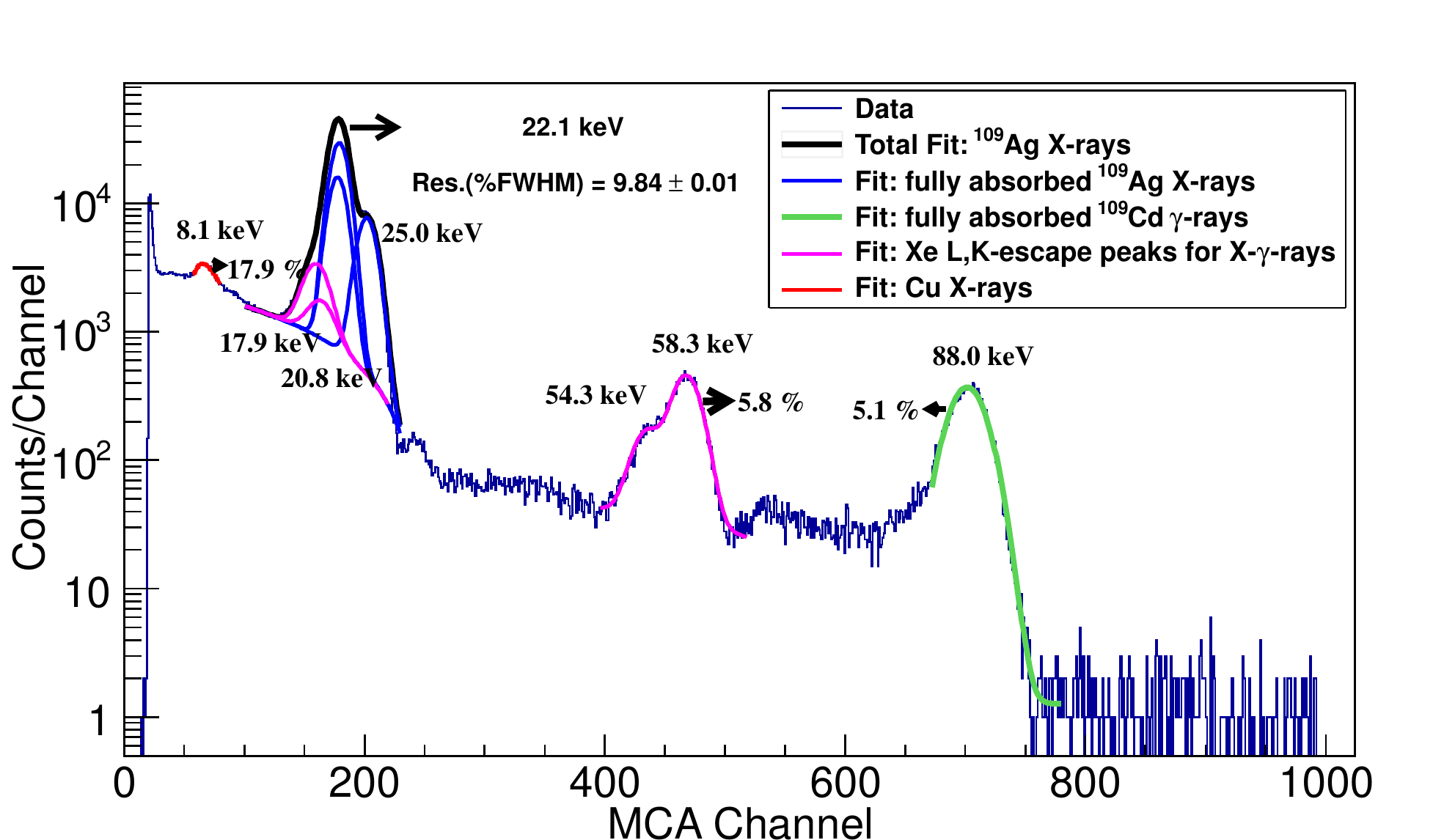}
\par\end{centering}
\caption{Energy spectrum from a $^{109}$Cd source acquired at 8 bar in a Xe+$1.4$\% TMA mixture, 
with $E/p=~245$ V/cm/bar. The K-fluorescence lines from $^{109}$Ag with their respective escape peaks of Xe are shown. 
The $\gamma$-line from $^{109}$Cd at $88.0$ keV is observed as well as the escape peak at $58.3$ keV. 
The fits realized for each peak and the energy resolution (\%FWHM) are depicted on the spectrum.
\label{fig:Energy-spectrum-at 8bar}}
\end{figure}

Figure \ref{fig:Energy-spectrum-at 8bar} shows an energy spectrum acquired at $8$ bar using a mixture of Xe+$1.4\%$ TMA, 
with a larger energy range than previous figure. It is interesting to observe the $\gamma$-rays
from $^{109}$Cd at $88.04$ keV and two escape peaks located at
$58.3$ keV and $54.3$ keV, which correspond to K$_{\alpha}$($29.7$
keV) and K$_{\beta}$ ($33.7$ keV) of Xe, respectively. 
The energy resolution of the main peaks is depicted, with values of 
$9.8$, $5.8$, $5.1$\,\% FWHM at $22.1$, $58.3$ and $88.0$ keV, respectively. 
As expected, the energy resolution shows a dependency with the inverse
of the square-root of the energy. We must point out that the energy spectrum in figure \ref{fig:Energy-spectrum-at 8bar} was not acquired under optimum 
fields for this mixture and the only interest was to observe the $\gamma$-rays of $^{109}$Cd.
The event containment and conversion probability of $^{109}$Cd $\gamma$-rays is indeed only well
suited for the highest pressures, making their study at low/medium presssure more complicated in our setup.

\section{Results and discussion}

The first goal of this work was to establish a range of TMA concentration in which it would
be possible to obtain the best energy resolution and highest gain for pressures between $1$ and $10\,\mbox{bar}$.
Hence, a systematic variation of the TMA concentrations was performed at four reference pressures: $1$,  
$5$, $8$ and $10\,\mbox{bar}$ and results are presented in section \ref{subsection:varTMA}.
An optimal range of TMA concentration was thus estimated, in which Penning-like transfer processes from excited Xe states are maximized.
Once the optimum concentration range was found, we performed a systematic study varying the pressure 
between $1$ and $10$ bar, which is presented in section \ref{subsection:VarPre}.

\subsection{Optimum concentration of TMA}\label{subsection:varTMA}

The variation in TMA concentration was performed at four pressures
in different ranges: at 1 bar ($0.4\,\%-15.5\,\%$), 2 bar ($0.4\,\%-6.0\,\%$),
at 8 bar ($0.3\,\%-5.0\,\%$), and 10 bar ($0.8\,\%-6.2\,\%$).
The active response of the filter to TMA led in practice to slightly different families of mixtures for each pressure.
The optimum range of TMA concentration was selected based on the dependences of the gain and the energy resolution with the amplification field, as described below.

\subsubsection*{4.1.1 Gas gain}

\begin{figure}[t]
\begin{centering}
\includegraphics[scale=0.38]{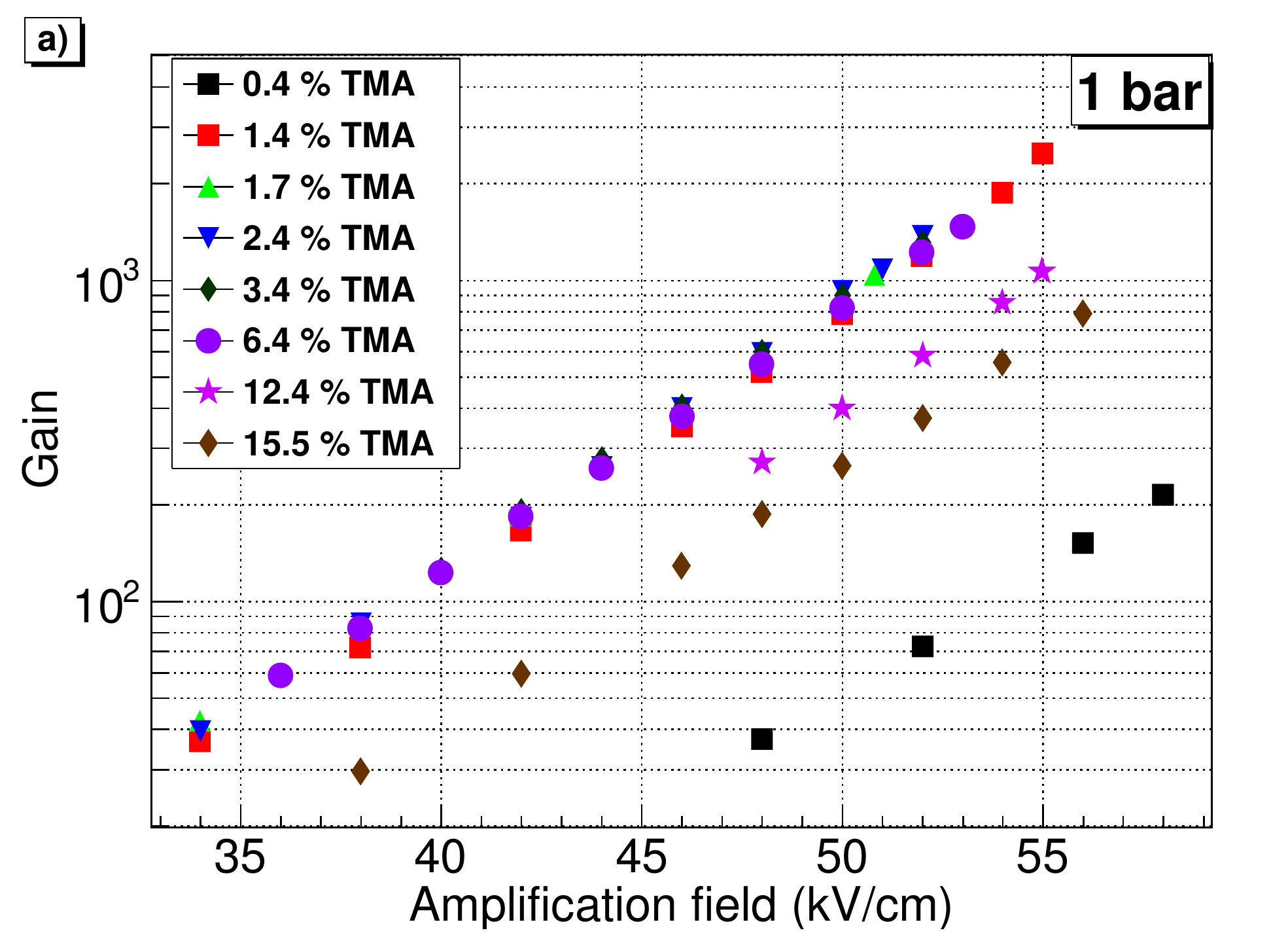}\includegraphics[scale=0.38]{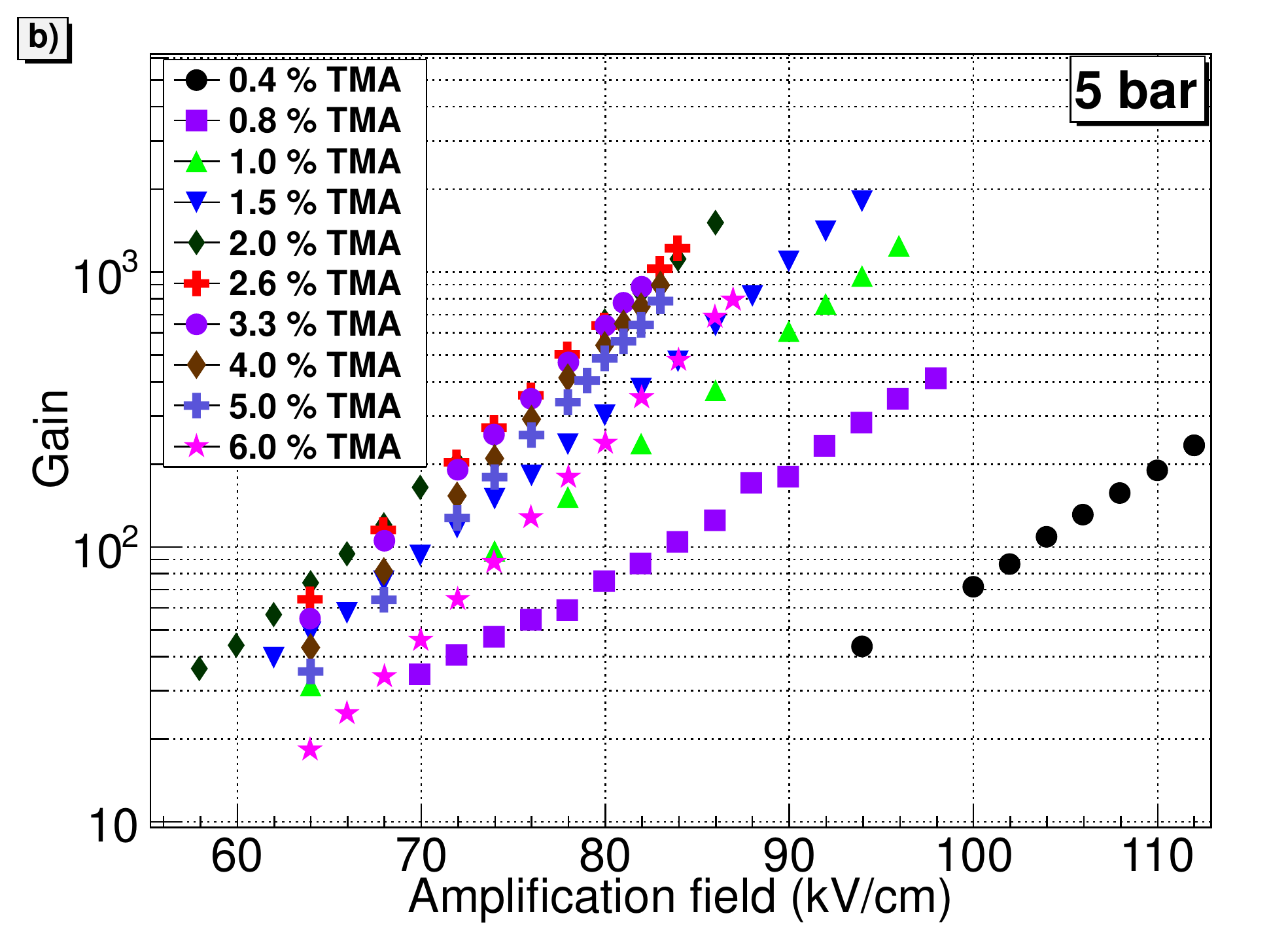}
\par\end{centering}
\begin{centering}
\includegraphics[scale=0.38]{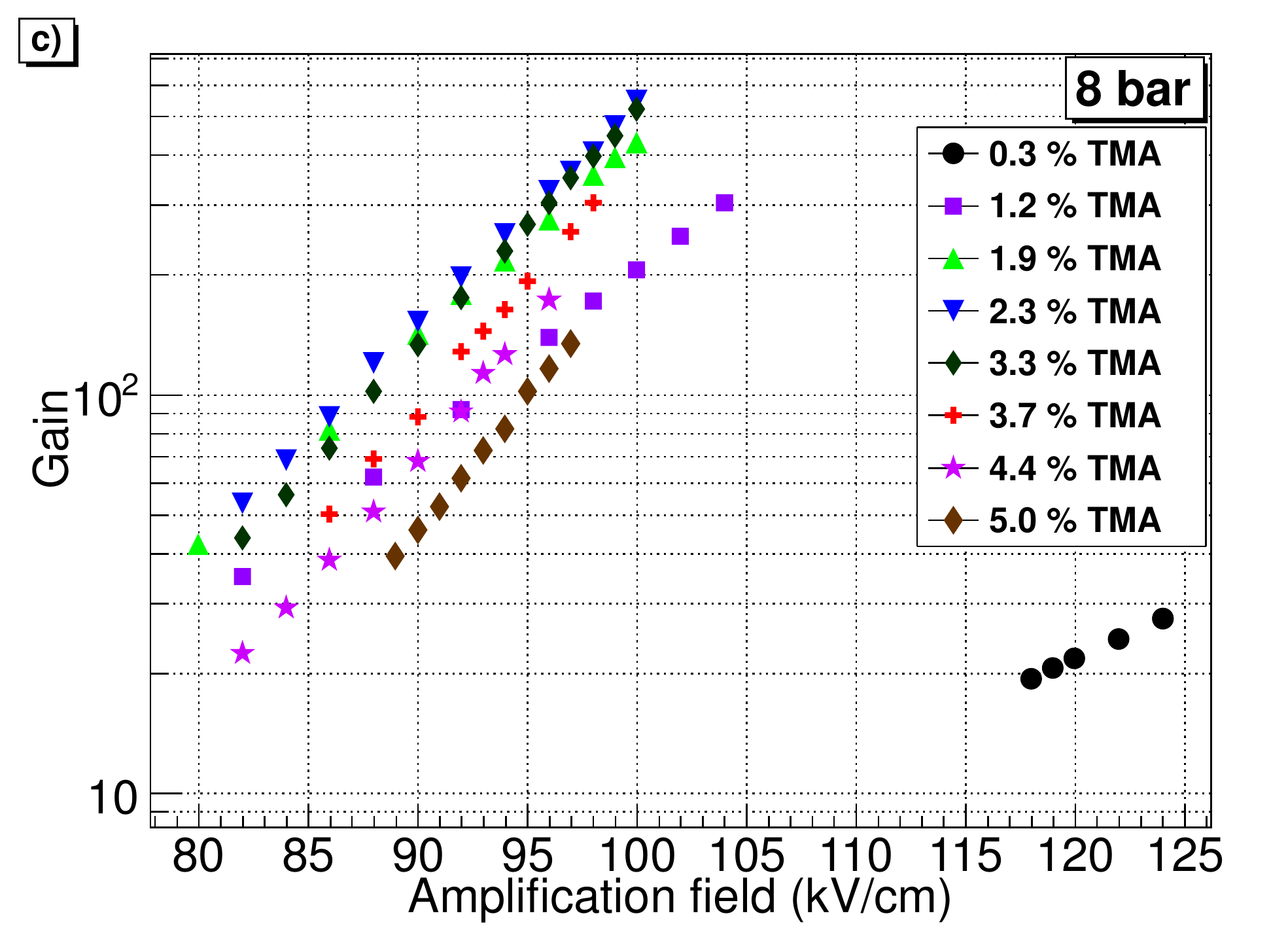}\includegraphics[scale=0.38]{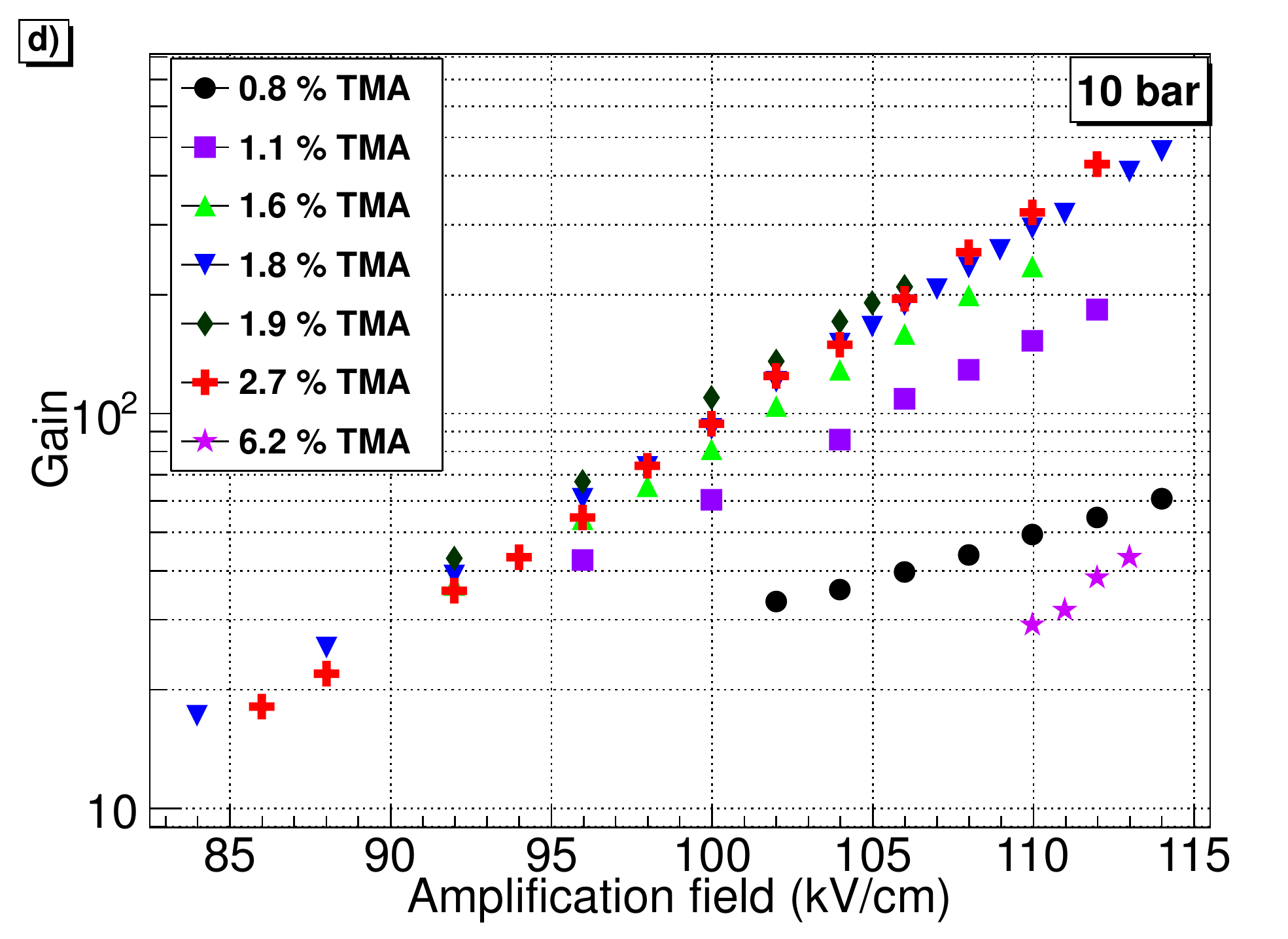}
\par\end{centering}
\caption{Dependency of gas gain with amplification field for different
TMA concentrations at 1 (a), 5 (b), 8 (c), 10 (d) bar. In each
graph the TMA concentration is indicated. \label{fig:GGainvsAmFi_Allpressures}}
\end{figure}

The dependence of the gain with the amplification field is shown in figure \ref{fig:GGainvsAmFi_Allpressures} at four pressures:
$1$ (a), $5$ (b) , $8$ (c), and $10$ (d) bar. 
The gain curves show a linear behaviour with the amplification field in the semi-log plot, having correlation factors above $0.999$
except for special cases with low statistics. 

Considering the plot at 1 bar (see figure \ref{fig:GGainvsAmFi_Allpressures}a), 
we observe that lower amplification fields must be applied
when the TMA percentage is increased from $0.4\%$ to $1.4$\%.  
The curves within a concentration range from $1.4\%$ and $6.4\%$ TMA seem to overlap, suggesting
that transfer mechanisms are already fully active while the avalanche dynamics remains largely unaffected.
The tendency reverses above $6.4\%$ TMA so that higher fields must be applied to obtain the same gas gain,
as also observed in neon-based mixtures \cite{PacoSaclay}.
A similar behaviour is observed at high pressures (see figure \ref{fig:GGainvsAmFi_Allpressures}b-d). 
This fact is more clear at high pressures, when low quenched mixtures are compared with highly quenched 
ones, being the trend also present at $1$ bar, but less pronounced.

\begin{figure}[hb]
\begin{centering}
\includegraphics[scale=0.4]{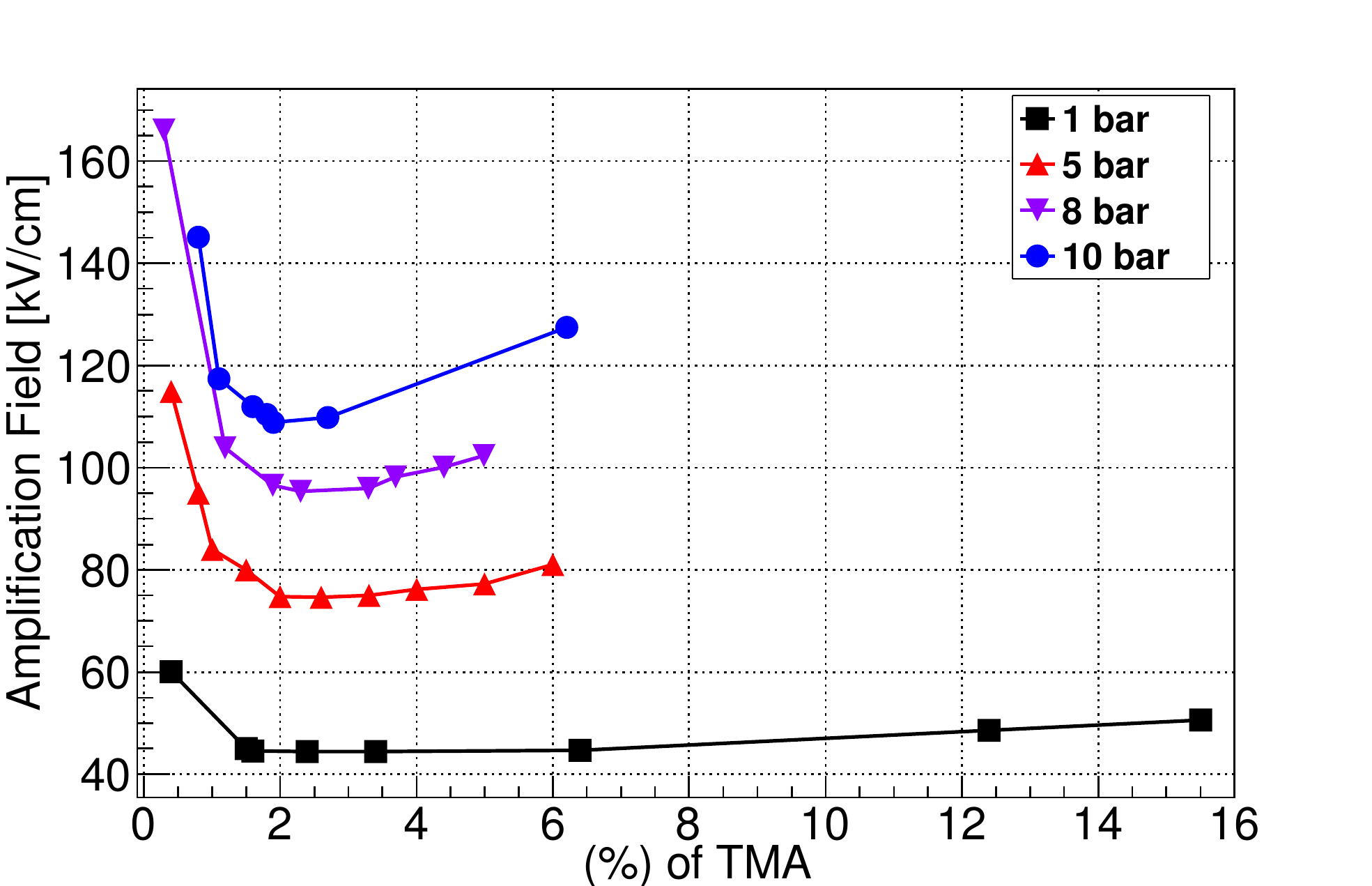}\includegraphics[scale=0.4]{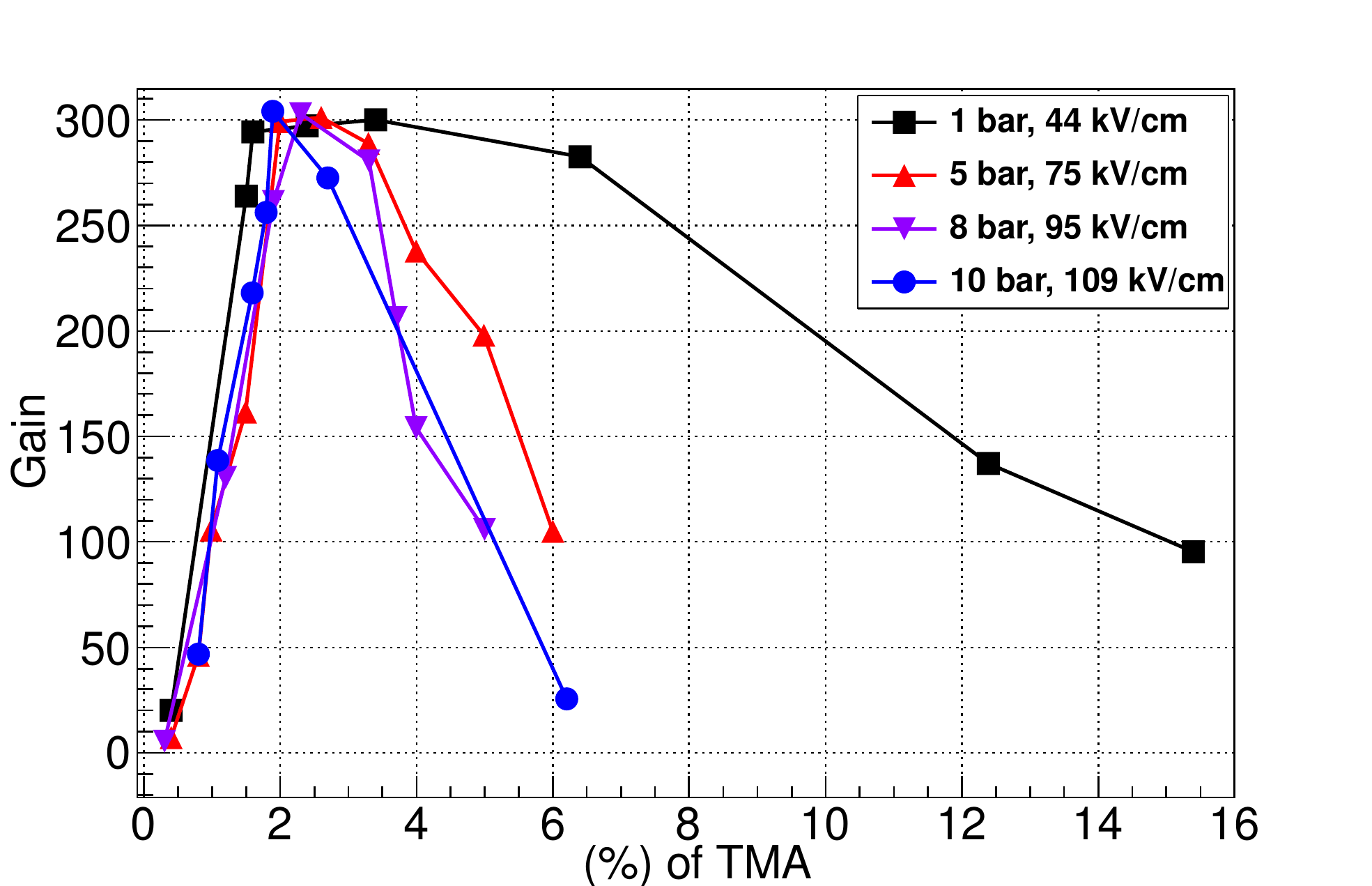}
\par\end{centering}
\caption{
Left: Amplification field needed for a gas gain of $G=300$ as a function of the TMA percentage.
Right: Gas gain as a function of the TMA percentage at a fixed pressure and amplification field. 
For each data set, an optimum for the the Penning effect can be clearly observed. \label{fig:AmFivsPTMA_1_5_8}}
\end{figure}

In order to do a better study of the optimum TMA concentration, 
we performed linear fits of the $\ln G$ versus amplification field data for proper interpolation or
extrapolation of the gain or field. For instance, figure \ref{fig:AmFivsPTMA_1_5_8}(left) shows 
the dependence of the amplification field calculated to obtain a gas gain of $300$ 
versus the TMA concentration at each pressure.
If we consider the points at $1$\,bar, by increasing the TMA concentration the field needed to obtain the same gain is reduced down to $44$ kV/cm for $1.7\,\%$ TMA. 
This field remains practically constant between $1.4\%$ and $6.4$\% TMA, increasing at higher concentrations. 
The same dependency is observed at high pressures: 
the amplification field reaches a minimum value at around $2\,\%$ TMA, being
 $75$, $95$, $109$ kV/cm at $5$, $8$ and $10$ bar, respectively.

On the other hand, representing the variation of the gas gain with the percentage of TMA is
another way to see that an optimum TMA concentration is reached at each pressure, 
for which the influence of transfer processes is maximized.
The minimum values of the amplification field reached before for gas gains of $300$, were used to 
study the behaviour of the gain (at each pressure) as a function of the TMA concentration (see figure \ref{fig:AmFivsPTMA_1_5_8}b)(right).
The gain rapidly increases when small quantities of TMA are added, increasing by a factor of $16$ ($50$) at $1$ ($5$) bar 
when TMA increases from $0.4\%$ to $2\%$.
The rapid rise in gas gain at constant amplification field has also been observed using Xe+$2$,$3$ dimethyl-2-butene as Penning additive, 
showing similar dependences with the additive at $1$ bar \cite{DMB}.  
Even stronger changes are seen at $8$ and $10$ bar. This great increase is a strong 
evidence that Penning effect takes place, being stronger at high pressure. 
In general, the gas gain reaches a maximum value within the range of $2\%$-$2.5\%$, then it drops.

The onset of breakdown at each pressure presents a similar behaviour: 
by increasing the TMA concentration, the maximum gas gain increases up to maximal values within a TMA concentration range of $2\%$ to $3\%$; 
above these values, the maximum gain falls. Values of maximum gas gain below $60$ are obtained
for TMA concentrations less than $1\%$, while within optimum range values gains above $2000$ ($400$) at $1$ ($10$) bar have been attained.    
This fact is an important advantage in comparison with pure Xe, because it permits to work 
at higher gas gain with lower amplification fields in stable conditions.
In some cases this trend was broken, attributed mainly to dust on the surface of the detector, 
which can limit the maximum gain, specially at high pressures.
Hence, the detector was routinely washed and then flushed with helium, 
reaching higher amplifications fields than before.


\subsubsection*{4.1.2 Energy Resolution}

\begin{figure}[h]
\begin{centering}
\includegraphics[scale=0.27]{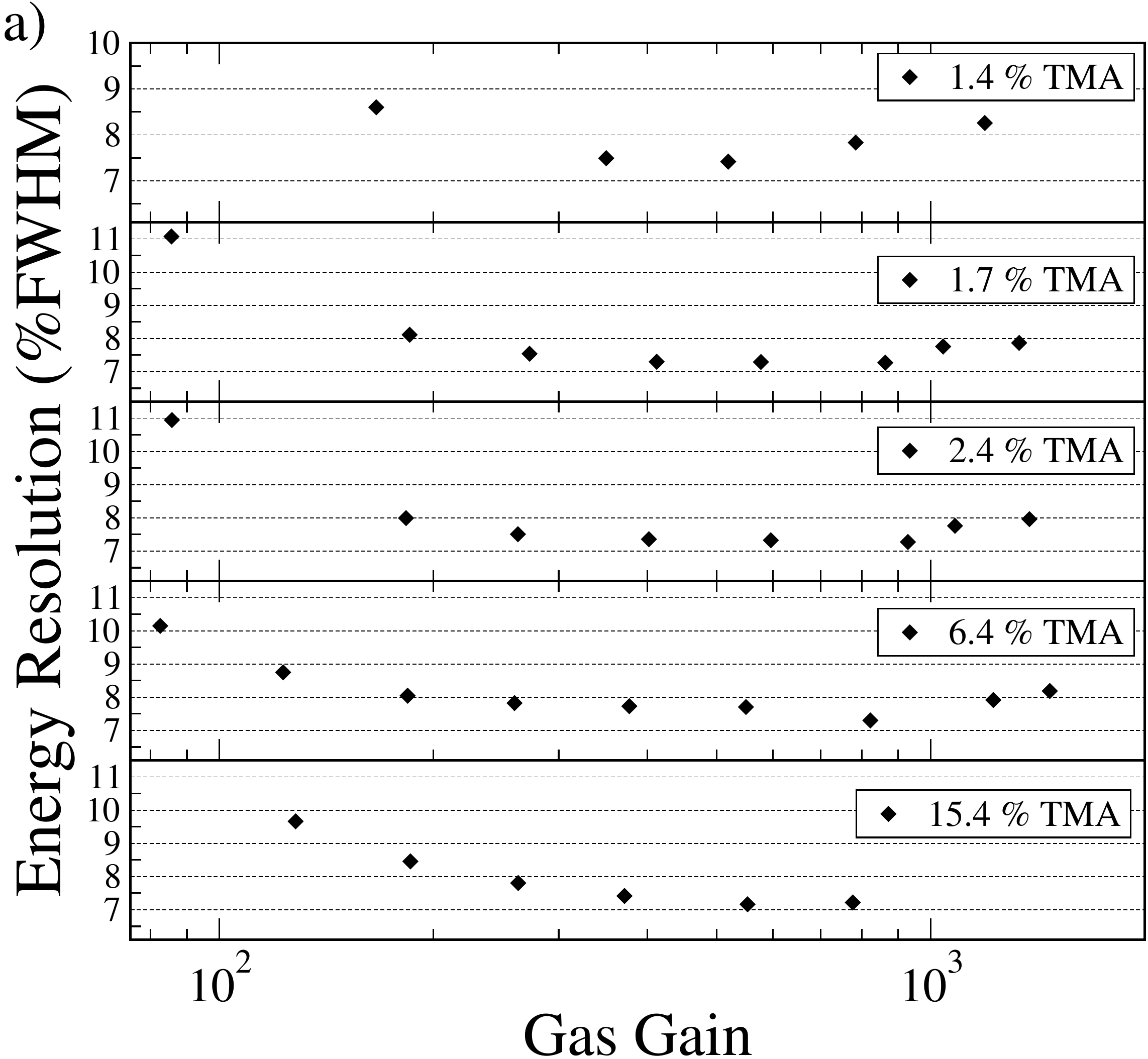}
~\includegraphics[scale=0.27]{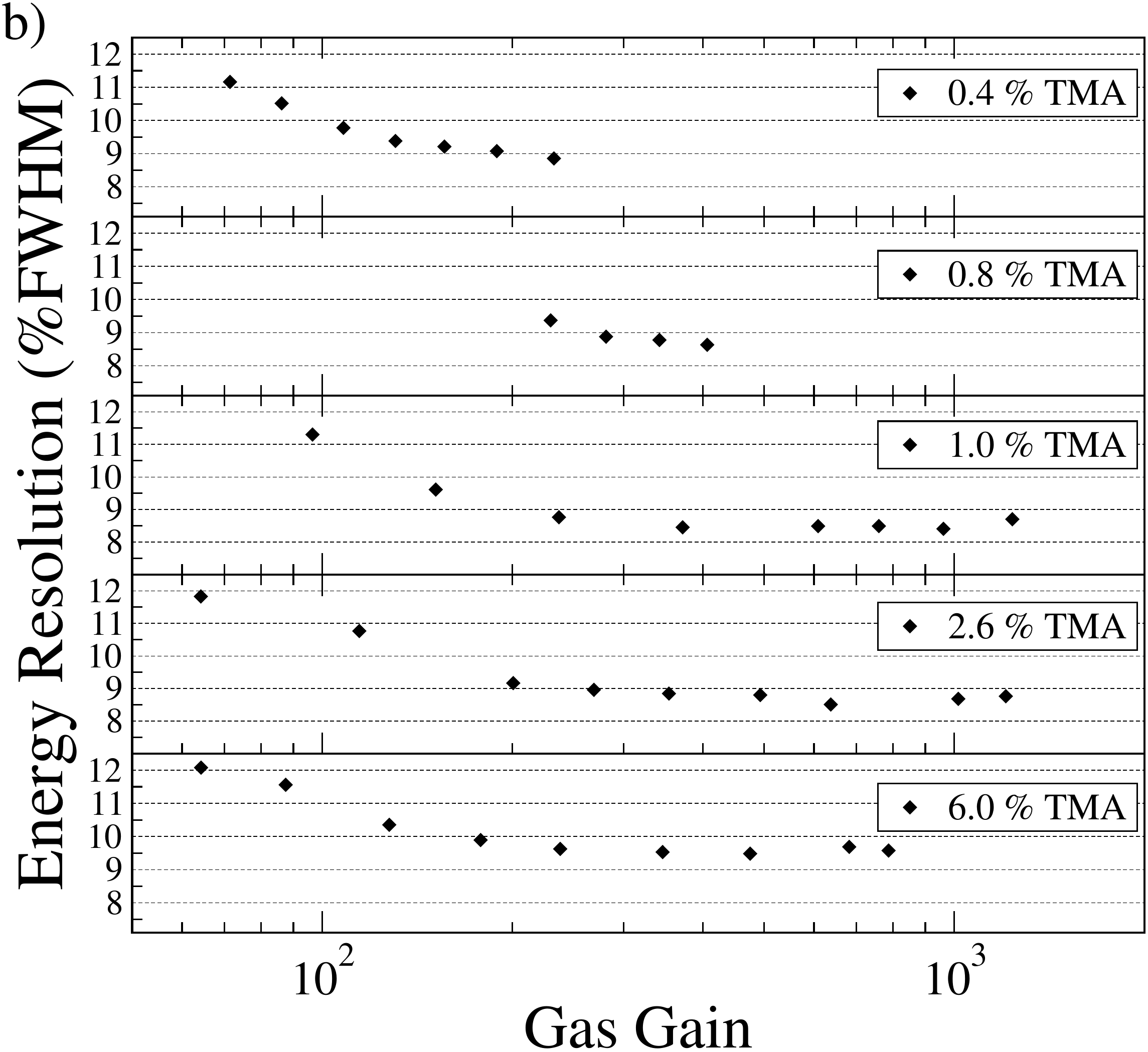}
\par\end{centering}
\begin{centering}
\includegraphics[scale=0.27]{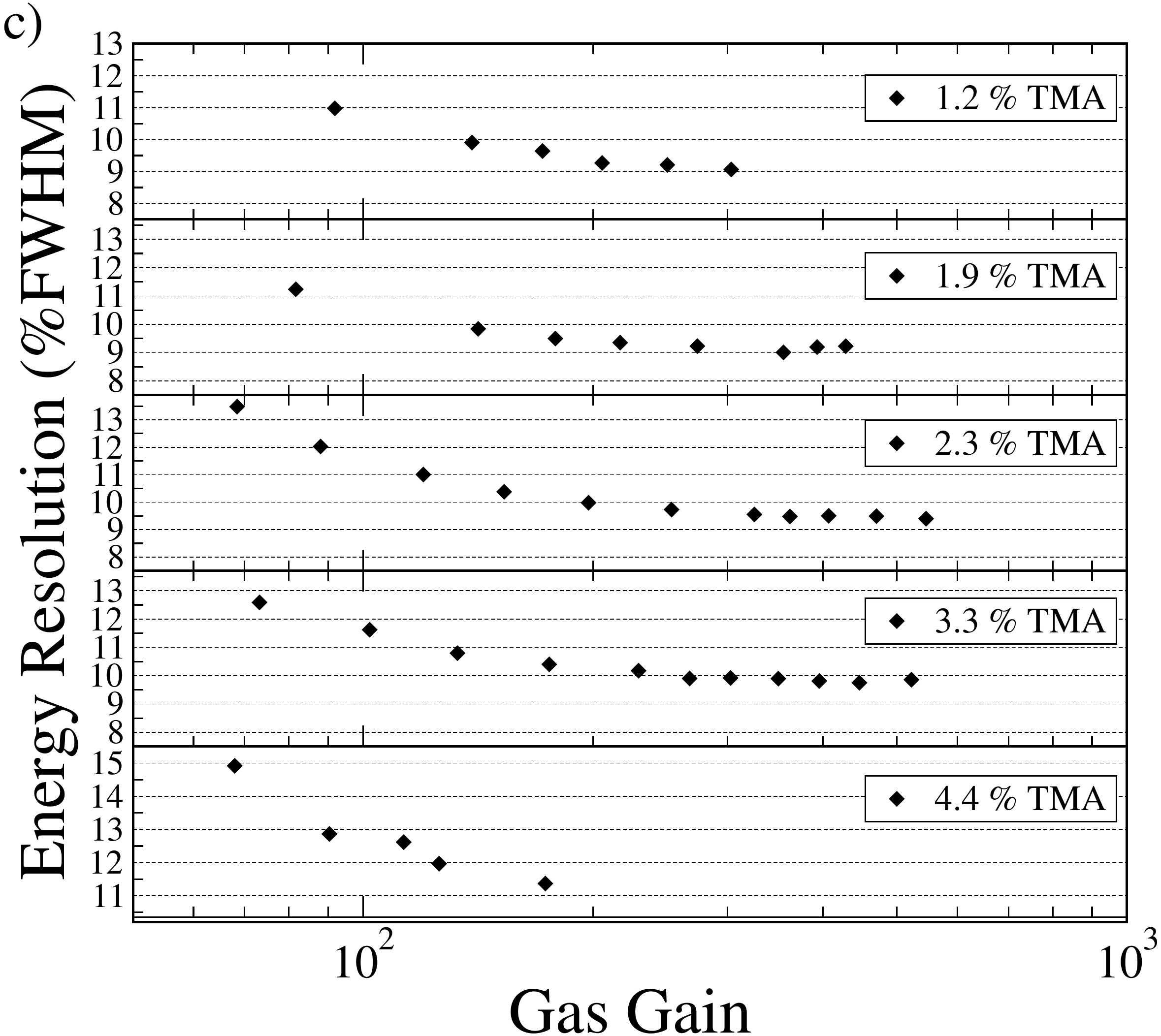}\includegraphics[scale=0.27]{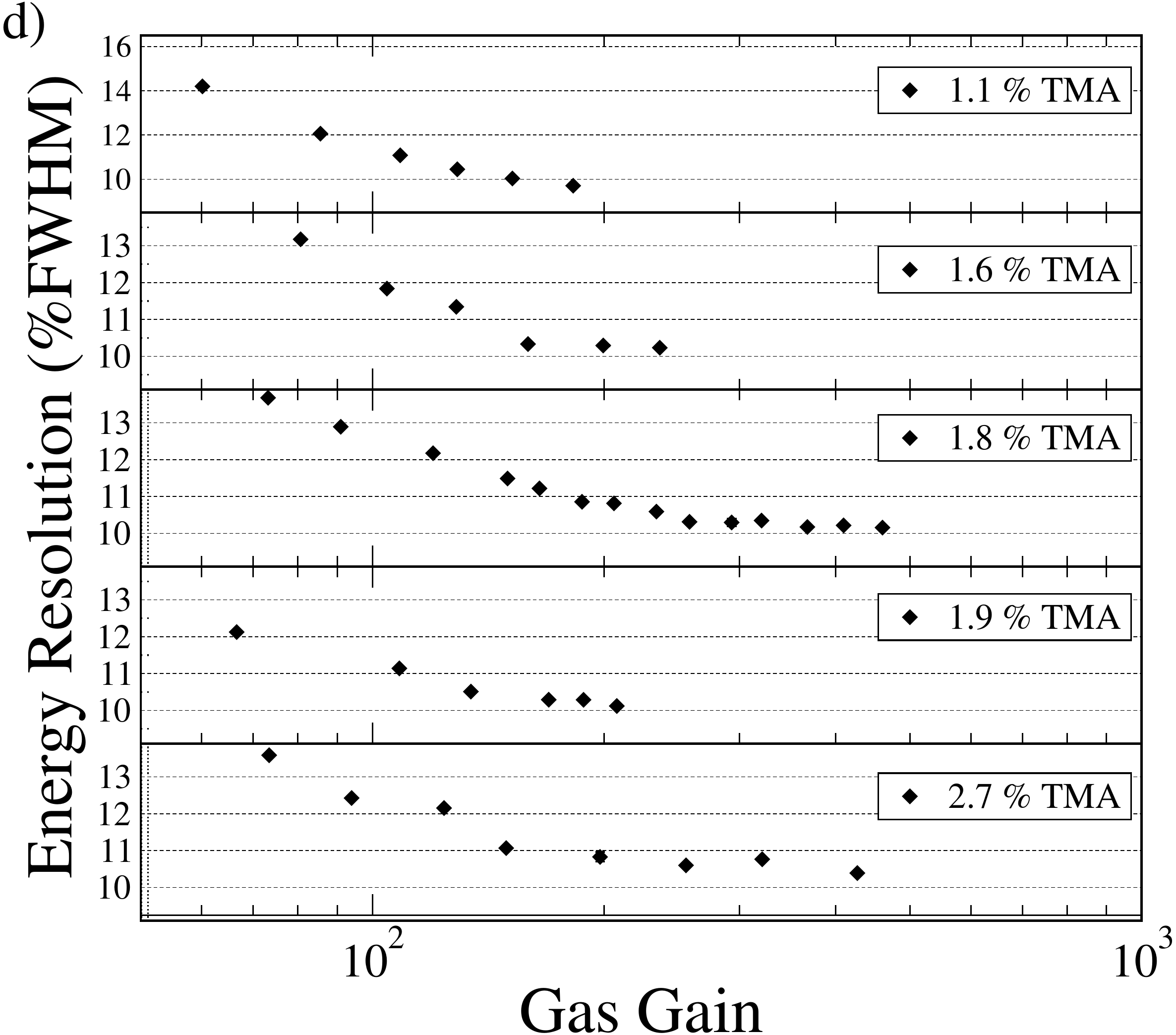}
\par\end{centering}
\centering{}\caption{Energy resolution at $22.1\,\mbox{keV}$ as a function of gas gain
for different TMA concentrations at 1 (a), 5 (b), 8 (c) and 10 bar (d). The TMA concentration is
indicated for each data set. \label{fig:EnRevsGGain_somepressurs} }
\end{figure}

The dependence of the energy resolution with the gas gain for  $22.1$ keV X-rays is shown in figure 
\ref{fig:EnRevsGGain_somepressurs} for four pressures: 1 (a), 5 (b), 8 (c) and 10 bar (d). 
The energy resolution improves as the gas gain increases for all mixtures.
The worse values at low gains are due to electronic noise, 
so its contribution was measured at FWHM injecting at the amplifier input an AC-coupled square pulse, and subtracting its value in quadrature.
It was found that the resulting energy resolution has
compatible values in the gas gain range between $50$ and $900$.
This allows to estimate that the noise contribution for gas gains above $300$ is less than $1\%$.
Hereafter, all values of resolution presented or plotted in this paper include the noise contribution.
For Xe+TMA mixtures within the optimal range for Penning transfer of $1.5\%$-$3\%$, 
the energy resolution degrades at the highest gas gain, 
more slowly at $5$ bar than at $1$ bar. 
However, this deterioration was not observed at high pressures ($8$ and $10$ bar) 
since the maximum gas gain was below $500$. 
Previous authors \cite{Xeup10bar} have suggested that this deterioration is caused by space charge effects, despite the small gains involved.

In table \ref{tab:Ta.BestEnRevsTMA_alpre} the best values obtained for the 
energy resolution are listed together with the gas gain at which they were measured.
At each pressure, the energy resolution improves adding TMA, taking minimum values around $2\%$ TMA and deteriorating at higher concentrations, 
At 1 bar, however, the minimum is not observed.
For instance, at $1$ bar, the energy resolution goes from a value of $8.4\%$ ($0.4\,\%$ TMA) 
down to $7.3\%$ FWHM at $22.1$ keV ($1.7\%$ TMA), 
then it remains constant between $1.4\%$ and $15.4\%$, for gains above $800$. 
These values of the energy resolution are compatible with measurements performed 
on a cylindrical proportional counter using $95\%$ Xe + $5\%$ TMA at $1$ bar \cite{XeBas}. 
On the other hand, at high pressures the energy resolution deteriorates
at even lower TMA percentages, the range where compatible values are obtained being $\sim1.0\%$- $2.6\%$. 
The best energy resolution is $8.3\%$, $9.0\,\%$ and $9.6\%$ FWHM 
for $5$, $8$ and $10$ bar, respectively. 
It must be noted, that we observe more variations in energy resolution within this optimum range, 
which are more pronounced with pressure. 
 We estimated our level of impurities below $30$ ppm for all mixtures, based on the 
relative variations determined with the mass spectrometer and considering that our values for the energy resolution are 
compatible with those quoted in \cite{Xeup10bar}.

As a summary, we can infer from these results that the best energy resolution can be obtained using 
TMA fractions between $1\%$ and $2.5\%$,
while in the previous section we have seen that the maximum gas gains are attained in a range 
of $2\%$ to $3\%$ TMA.  
Therefore, we can conclude that optimal Xe+TMA mixtures should be ranging from $1.5\%$ to $2.5\%$.
For applications where the energy resolution is the most important factor, 
TMA concentrations above $1\%$ should be used.
\begin{table}[H]
\begin{centering}
{\footnotesize }\begin{tabular}{c|c|c|c|c}
\hline 
\multicolumn{1}{c|}{{\small \footnotesize \% TMA}} & {\small \footnotesize Best Ene. Res. (\% FWHM)} & {\small \footnotesize Gain {[}$\times10^{2}${]} } & $A$ (cm$^{-1}$bar$^{-1}$) & $B$ (kV/cm/bar)\tabularnewline
\hline 
\multicolumn{3}{l}{{\footnotesize Pressure 1 bar}} & \multicolumn{1}{c}{}   \tabularnewline
\hline
{\small \footnotesize $0.4$} & {\footnotesize $8.4$} & {\footnotesize $1.5$} & {\footnotesize 6986$\pm$284} & {\footnotesize 108.9$\pm$2.0} \tabularnewline
{\footnotesize $1.4$} & {\footnotesize $7.4$} & {\footnotesize $5.2$} & {\footnotesize 5007$\pm$251} & {\footnotesize 66.7$\pm$2.1} \tabularnewline
{\footnotesize $1.7$} & {\footnotesize $7.3$} & {\footnotesize $8.6$} & {\footnotesize 4507$\pm$96} & {\footnotesize 61.1$\pm$0.8} \tabularnewline
{\footnotesize $2.4$} & {\footnotesize $7.3$} & {\footnotesize $9.3$} & {\footnotesize 5073$\pm$167} & {\footnotesize 66.3$\pm$1.4} \tabularnewline
{\footnotesize $3.4$} & \footnotesize $7.5$ &\footnotesize $9.0$& {\footnotesize 5058$\pm$172} & {\footnotesize 66.1$\pm$1.4} \tabularnewline
{\footnotesize $6.4$} & {\footnotesize $7.3$} & {\footnotesize $8.2$} & {\footnotesize 4974$\pm$153} & {\footnotesize 65.7$\pm$1.3} \tabularnewline
{\footnotesize $12.4$} & {\footnotesize $7.3$} & {\footnotesize $5.8$} & {\footnotesize 6020$\pm$672} & {\footnotesize 80.8$\pm$5.6} \tabularnewline
{\footnotesize $15.5$} & {\footnotesize $7.3$} & {\footnotesize $5.5$} & {\footnotesize 5714$\pm$304} & {\footnotesize 81.7$\pm$2.5} \tabularnewline
\hline
\multicolumn{3}{l}{{\footnotesize Pressure 5 bar}} & \multicolumn{1}{c}{}  & \tabularnewline
\hline
{\footnotesize $0.4$} & {\footnotesize $8.9$} & {\footnotesize $2.3$} & {\footnotesize 1665$\pm$81} & {\footnotesize 45.5$\pm$1.0}\tabularnewline
{\footnotesize $0.8$} & {\footnotesize $8.6$} & {\footnotesize $4.1$} & {\footnotesize 988$\pm$16} & {\footnotesize 27.9$\pm$0.3} \tabularnewline
{\footnotesize $1.0$} & {\footnotesize $8.4$} & {\footnotesize $7.3$} & {\footnotesize 1182$\pm$46} & {\footnotesize 27.6$\pm$0.6}\tabularnewline
{\footnotesize $1.5$} & {\footnotesize $8.3$} & {\footnotesize $8.1$} & {\footnotesize 1144$\pm$46} & {\footnotesize 25.8$\pm$0.6} \tabularnewline
{\footnotesize $2.0$} & {\footnotesize $8.4$} & {\footnotesize 6.5} & {\footnotesize 1228$\pm$35} & {\footnotesize 25.2$\pm$0.4} \tabularnewline
{\footnotesize $2.6$} & {\footnotesize $8.5$} & {\footnotesize 6.4} & {\footnotesize 1492$\pm$23} & {\footnotesize 28.1$\pm$0.2} \tabularnewline
{\footnotesize $3.3$} & {\footnotesize $8.9$} & {\footnotesize 8.8} & {\footnotesize 1772$\pm$39} & {\footnotesize 30.7$\pm$0.3}\tabularnewline
{\footnotesize $4.0$} & {\footnotesize $9.0$} & {\footnotesize 7.4} & {\footnotesize 2008$\pm$38} & {\footnotesize 33.1$\pm$0.3} \tabularnewline
{\footnotesize $5.0$} & {\footnotesize $9.0$} & {\footnotesize 7.8} & {\footnotesize 2246$\pm$43} & {\footnotesize 35.3$\pm$0.3}\tabularnewline
{\footnotesize $6.0$} & {\footnotesize $9.6$} & {\footnotesize 7.9} & {\footnotesize 2783$\pm$15} & {\footnotesize 40.6$\pm$0.1}\tabularnewline
\hline
\multicolumn{3}{l}{{\footnotesize Pressure 8 bar}} & \multicolumn{1}{c}{} & \tabularnewline
\hline
{\footnotesize 0.3} & {\footnotesize 27.3} & {\footnotesize 0.2} & {\footnotesize 708$\pm$79} & {\footnotesize 33.3$\pm$1.7} \tabularnewline
{\footnotesize 1.2} & {\footnotesize 9.1} & {\footnotesize 3.0} & {\footnotesize 809$\pm$58} & {\footnotesize 22.7$\pm$0.8} \tabularnewline
{\footnotesize 1.9} & {\footnotesize 9.0} & {\footnotesize 3.6} & {\footnotesize 1162$\pm$99} & {\footnotesize 25.3$\pm$1.0}\tabularnewline
{\footnotesize 2.3} & {\footnotesize 9.4} & {\footnotesize 5.4} & {\footnotesize 1379$\pm$42} & {\footnotesize 27.0$\pm$0.3} \tabularnewline
{\footnotesize 3.3} & {\footnotesize 9.8} & {\footnotesize 4.5} & {\footnotesize 1696$\pm$41} & {\footnotesize 29.6$\pm$0.3} \tabularnewline
{\footnotesize 3.7} & {\footnotesize 10.4} & {\footnotesize 3.0} & {\footnotesize 2224$\pm$90} & {\footnotesize 33.6$\pm$0.5} \tabularnewline
{\footnotesize 4.4} & {\footnotesize 11.4} & {\footnotesize 1.7} & {\footnotesize 2519$\pm$59} & {\footnotesize 35.7$\pm$0.3} \tabularnewline
{\footnotesize 5.0} & {\footnotesize 12.0} & {\footnotesize 1.3} & {\footnotesize 3399$\pm$222} & {\footnotesize 40.2$\pm$0.8} \tabularnewline
\hline 
\multicolumn{3}{l}{{\footnotesize Pressure 10 bar}} & \multicolumn{1}{c}{} & \tabularnewline
\hline 
{\footnotesize 0.8} & {\footnotesize 19.7} & {\footnotesize 0.6} & {\footnotesize 330$\pm$25} & {\footnotesize 15.9$\pm$0.8}\tabularnewline
{\footnotesize 1.1} & {\footnotesize 9.6} & {\footnotesize 1.8} & {\footnotesize 772$\pm$43} & {\footnotesize 22.4$\pm$0.6}\tabularnewline
{\footnotesize 1.6} & {\footnotesize 10.2} & {\footnotesize 2.3} & {\footnotesize 997$\pm$45} & {\footnotesize 24.3$\pm$0.5} \tabularnewline
{\footnotesize 1.8} & {\footnotesize 10.2} & {\footnotesize 4.6} & {\footnotesize 1072$\pm$13} & {\footnotesize 24.7$\pm$0.1} \tabularnewline
{\footnotesize 1.9} & {\footnotesize 10.1} & {\footnotesize 2.0} & {\footnotesize 1141$\pm$54} & {\footnotesize 25.0$\pm$0.5}\tabularnewline
{\footnotesize 2.7} & {\footnotesize 10.4} & {\footnotesize 4.2} & {\footnotesize 1423$\pm$29} & {\footnotesize 27.6$\pm$0.2} \tabularnewline
{\footnotesize 6.2} &  {\footnotesize 47.0} & {\footnotesize 0.4} & {\footnotesize 4644$\pm$1033} & {\footnotesize 46.6$\pm$4.7}\tabularnewline
\hline
\end{tabular}
\par\end{centering}{\footnotesize \par}

\caption{Values of the best energy resolution obtained at each TMA concentration
for pressures of $1$, $5$, $8$ and $10\,\mbox{bar}$, with the corresponding
value of gas gain at which it was obtained. The systematic error of all values is estimated to be less than $0.2\%$ FWHM.
Values of the parameters $A$ and $B$ together with their errors are obtained from 
the fit to the equation $4.1$. \label{tab:Ta.BestEnRevsTMA_alpre}}
\end{table}

\subsubsection*{4.1.3 First Townsend parameter}

The first Townsend coefficient\, ($\alpha$)\, permits to\, describe the electron multiplication\, processes in\, Micromegas.
Taking into account a\, semi-empirical parameterization of this coefficient\, \cite{Sauli},\,\,
\begin{equation}
 \alpha~=~AP~\exp(-B/S) \label{eq:first}
\end{equation}
where $S$ is the reduced amplification field ($S=E_{amp}/P$),
$P$ is the gas pressure, and $A$ and $B$ are constants that depend on the gas.  
The logarithm of the gas gain of a Micromegas is the integral of the Townsend coefficient along the field lines from the mesh to the anode. 
By assuming a constant field, in first order, the electron multiplication $G$ can be written as \cite{XeMiTow}:
\begin{equation}
\ln\left(\frac{\ln(G)}{P}\right)=\ln(Ad)-\frac{B}{S}\label{eq:Main}
\end{equation}
where $d$ is mesh-anode distance and $A$ and $B$ can be obtained from a linear fit of the $\ln(\ln(G))$ vs. $S^{-1}$ data.
According to this simplified physical picture quoted by the authors of \cite{Xeup10bar,XeMethane}, 
$A$ and $B$ are related with microscopic parameters where:  
$A$ corresponds to the inverse of the collision mean free path at the reference pressure $P_0$ ($A=1/(\lambda_0P_0)$),
while  $B=V_i/(\lambda_0P_0)$, where $V_i$ is the effective energy 
to produce an electron-ion pair.

In table \ref{tab:Ta.BestEnRevsTMA_alpre} 
the corresponding parameters for each set of data are shown.
Under the physical picture in \cite{Xeup10bar}, the mean energy of the electron swarm between two collisions is given by
$\bar{\epsilon}=~E\lambda~=~E\lambda_0P_0/P$. 
The ratio $\bar{\epsilon}/V_i=E\lambda/V_i$ gives hence a figure on how likely ionization is to occur.
Naturally, for large $\lambda/V_i$, ionization will occur at lower fields. 
This implies that $B\times P= V_i/\lambda$ can be seen as a characteristic reduced field at which 
the probability of ionization becomes important. 
Therefore, the following interpretation of the observations can be made:

\begin{itemize}
\item[--] for small TMA concentrations $B$ starts from a given value, asymptotically approaching the one in pure Xenon.

\item[--] as TMA increases, Penning transfers become active, allowing for ionization to take place at lower fields. This
process is mediated by excited atomic and molecular states, that require considerably less energy
for being ionized than the parent gas, through transfer reactions. Therefore B is reduced. 
The measurable consequence of this fact is that the gain increases with TMA 
concentration by factors up to $\times 16-50$ at $1-10$ bar, with the electric field remaining unchanged.
\item[--] when TMA further increases, B becomes larger, probably caused by the TMA molecule cooling down the electron swarm.
Thus larger fields are required to achieve identical gas gains.
\end{itemize}

We conclude that, in Xe+TMA mixtures, a narrow operating range exists, where TMA is clearly advantageous over pure Xe for this
kind of amplification structures. This occurs within a TMA concentration range of $\sim1.4\%$-$6.4\%$ and $\sim1.4\%$-$2.5\%$ at 
$1$ and from $5$ to $10$ bar, respectively. 
It is known that quenched gases show a significantly reduced diffusion, improving the pattern recognition of pixelized readouts, as studied in \cite{LauraPaco}.


%




\subsection{Varying the pressure}\label{subsection:VarPre}

In this section we present the results of gas gain and energy resolution 
obtained when the pressure was varied from $1$ to $10$ bar, 
using TMA concentrations within the range that was estimated as optimum in the previous section. 
Therefore, mixtures with TMA concentrations ranging from $1.5\,\%$ to $2\,\%$ were selected, 
and the values are specified in table \ref{tab:mixtures}. 

\begin{figure}[hb]
\begin{centering}
\includegraphics[scale=0.4]{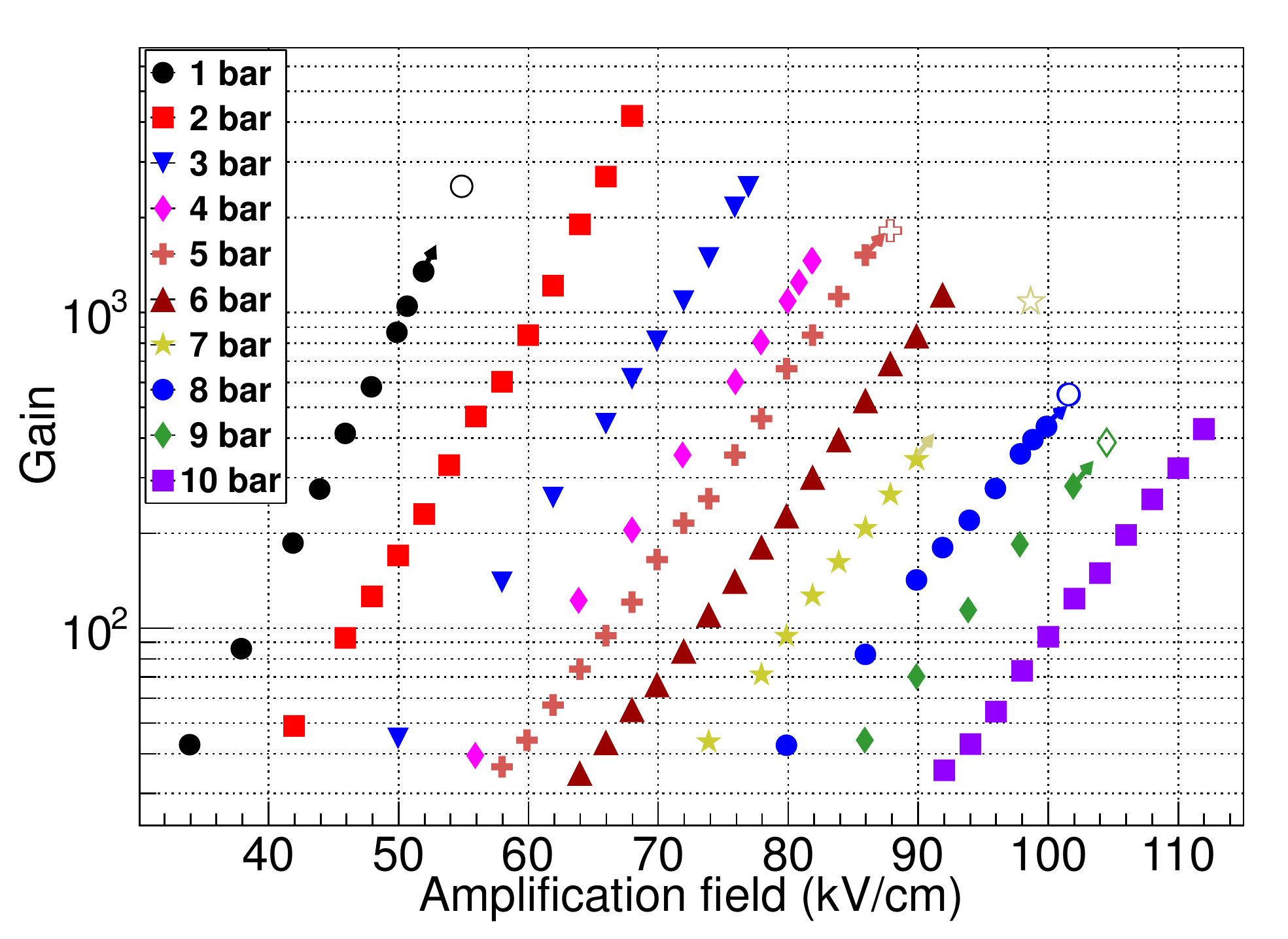}\includegraphics[scale=0.4]{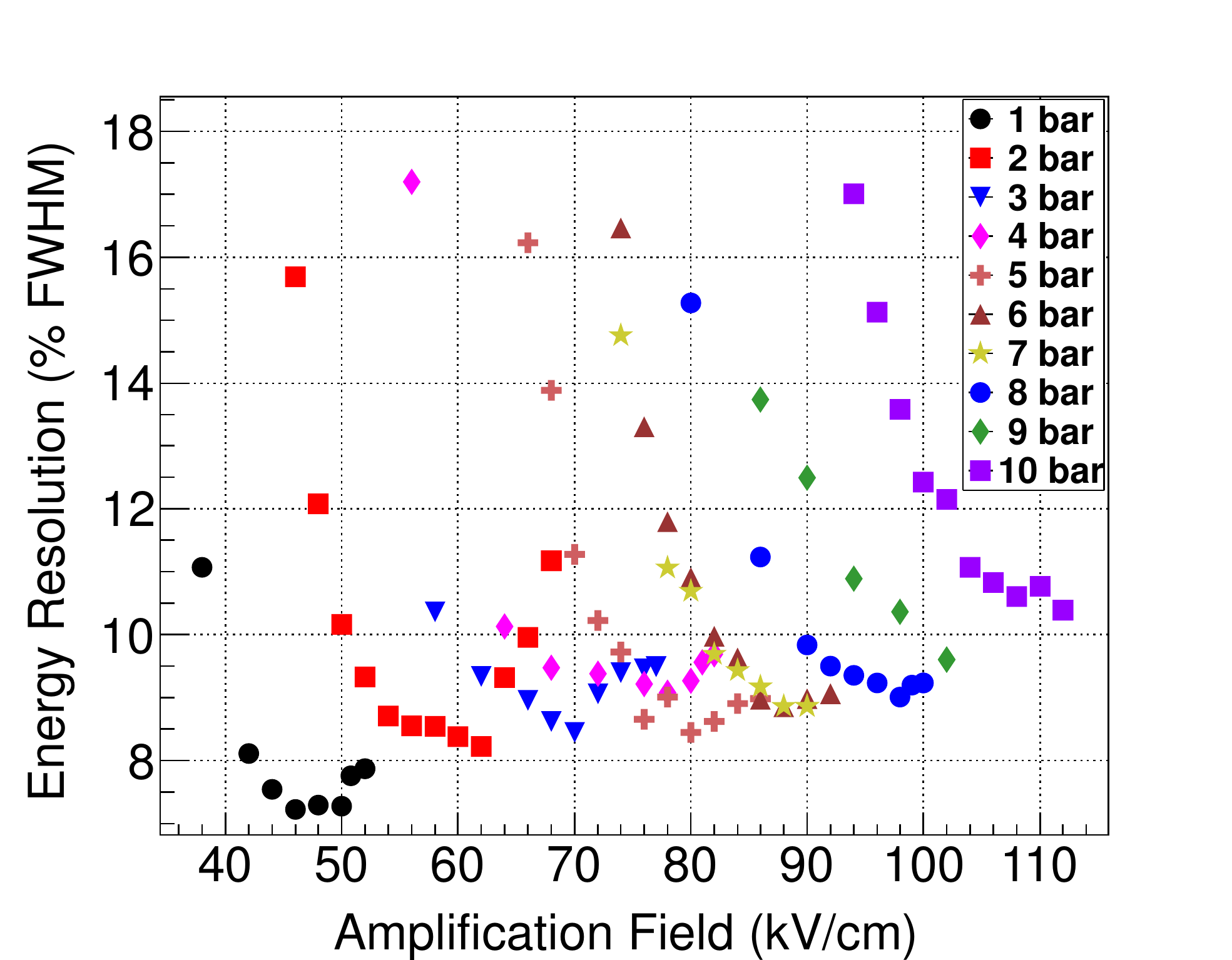}
\par\end{centering}
\caption{Dependence of the gain (left) and the energy resolution at $22.1$ keV (right) with 
the amplification field for a pressure ranging from $1$ to $10$ bar, at optimal TMA concentrations 
within the range $1.5\%$-$2\%$. 
An empty marker represents the maximum gain in case  it was reached in a different mixture 
but with TMA concentration close to optimum values.
\label{fig:GGainvsAmFi}}
\end{figure}

\begin{table}[H]
\begin{centering}
\begin{tabular}{c|c|c|c|c|c|c|c|c|c|c}
Pressure & 1 & 2 & 3 & 4 & 5 & 6 & 7 & 8 & 9 & 10\tabularnewline
\hline 
\% TMA & 1.7 & 1.7 & 1.5 & 1.6 & 2.0 & 2.0 & 2.0 & 1.9 & 1.9 & 1.8\tabularnewline
\end{tabular}
\par\end{centering}{\footnotesize \par}
\caption{TMA concentrations used for the systematic gas gain and energy resolution 
measurements when pressure was varied between $1$ and $10$ bar. 
Each TMA concentration is within the range that was considered as optimum for Penning transfer. \label{tab:mixtures}}
\end{table}
Gas gain curves are shown in figure \ref{fig:GGainvsAmFi} (left). 
It is noted that the amplification field necessary to reach any given gain increases with pressure,
as already observed in pure Xe \cite{Coimbra,Xeup10bar}. 
The slope tends to decrease with pressure, thus at high pressures for a given change in the amplification field, the absolute 
change in gain is lower according to what has been observed in pure Xe and Xe+methane mixtures \cite{Xeup10bar}.   
On the other hand, the maximum gain drops nearly exponentially with pressure for pressures above $2$ bar, 
down to $\sim400$ at $10$ bar. However, the maximum gain at any pressure is still at least 
a factor 3 higher than for Micromegas operated in pure Xe \cite{Coimbra}.
This behaviour is probably caused by photons in the avalanche generated from molecular Xe excitations \cite{Xeup10bar}.   


\begin{figure}[h]
\begin{centering}
\includegraphics[scale=0.4]{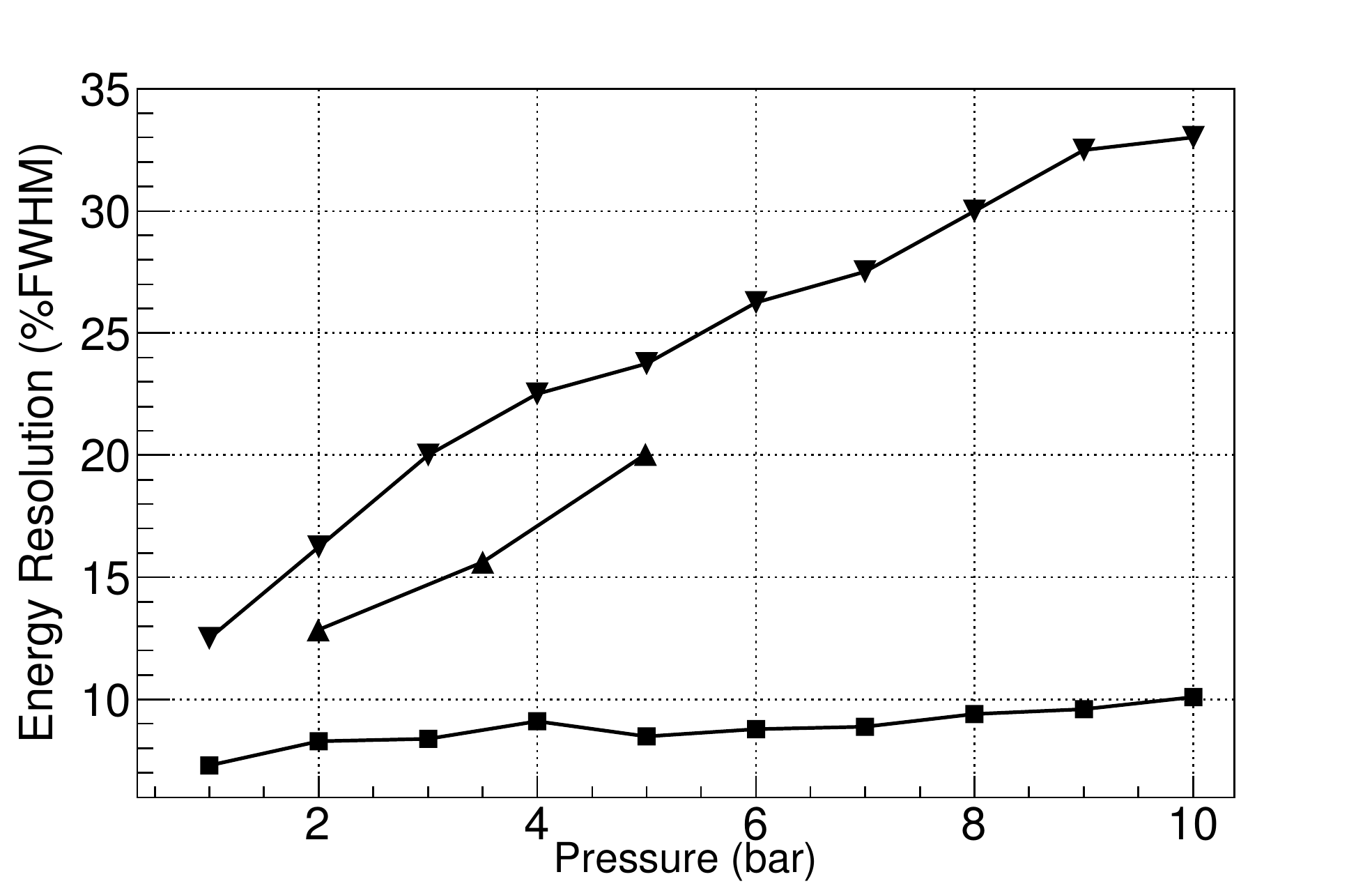}
\par\end{centering}

\caption{The dependence of energy resolution with pressure at 22.1 keV, for data from \cite{Coimbra} ($\blacktriangledown$), 
 data \cite{XeZaragoza} ($\blacktriangle$) and in this work ($\blacksquare$).
\label{fig:EnRevsPress}}
\end{figure}

The energy resolution (Figure \ref{fig:GGainvsAmFi}, right) shows a rapid improvement with 
the amplification field, reaching a minimum value and then degrading at high fields. 
As mentioned before, the signal-to-noise ratio explains the high values at low fields, 
The best energy resolution measured at each pressure for gains within the range from $300$ to $800$ is 
shown in figure \ref{fig:EnRevsPress} ($\blacksquare$).
A slight degradation with pressure is observed, 
which may be caused by electronegative impurities or inherent physical mechanisms \cite{Xeup10bar}.
Moreover, our deterioration in energy resolution with pressure has a tendency compatible with measurements
performed by H. Sakurai et al \cite{Xeup10bar},
were the estimated level of impurities was below $3$ ppm.
Thus we may conclude that a mechanism inherent to the avalanche process in HP Xe+TMA mixtures
is responsible for the observed degradation. 
As it has been noted, at higher pressure the reduced amplification field is weaker, 
a fact resulting in the increase of the number of excitations as compared to ionizations, 
hence more avalanche fluctuations and the subsequent degradation in the energy resolution \cite{Xeup10bar, XeMethane}.
We plan to address this statement more quantitatively in future works.

The results of this work are compared in figure \ref{fig:EnRevsPress} 
with previous measurements of Micromegas detectors in pure Xe:
with the same setup used in this study ($\blacktriangle$) \cite{XeZaragoza} 
and with a different one ($\blacktriangledown$) \cite{Coimbra}. 
The energy resolution achieved at $22.1$ keV is substantially better in this work, 
going down to $7.3\,\%$ ($9.6\,\%$) FWHM at $1$ ($10$) bar. 
This fact translates into an improvement of a factor $2$ ($3$) at $1$ ($10$) bar as compared to previous
measurements in pure Xe. 
Therefore, we can infer that the addition of TMA to Xe reduces the avalanche fluctuations;
likely due to transfer reactions from the Xe excited states to TMA molecules (Penning effect). 
\begin{figure}[h]
\begin{centering}
\includegraphics[scale=0.4]{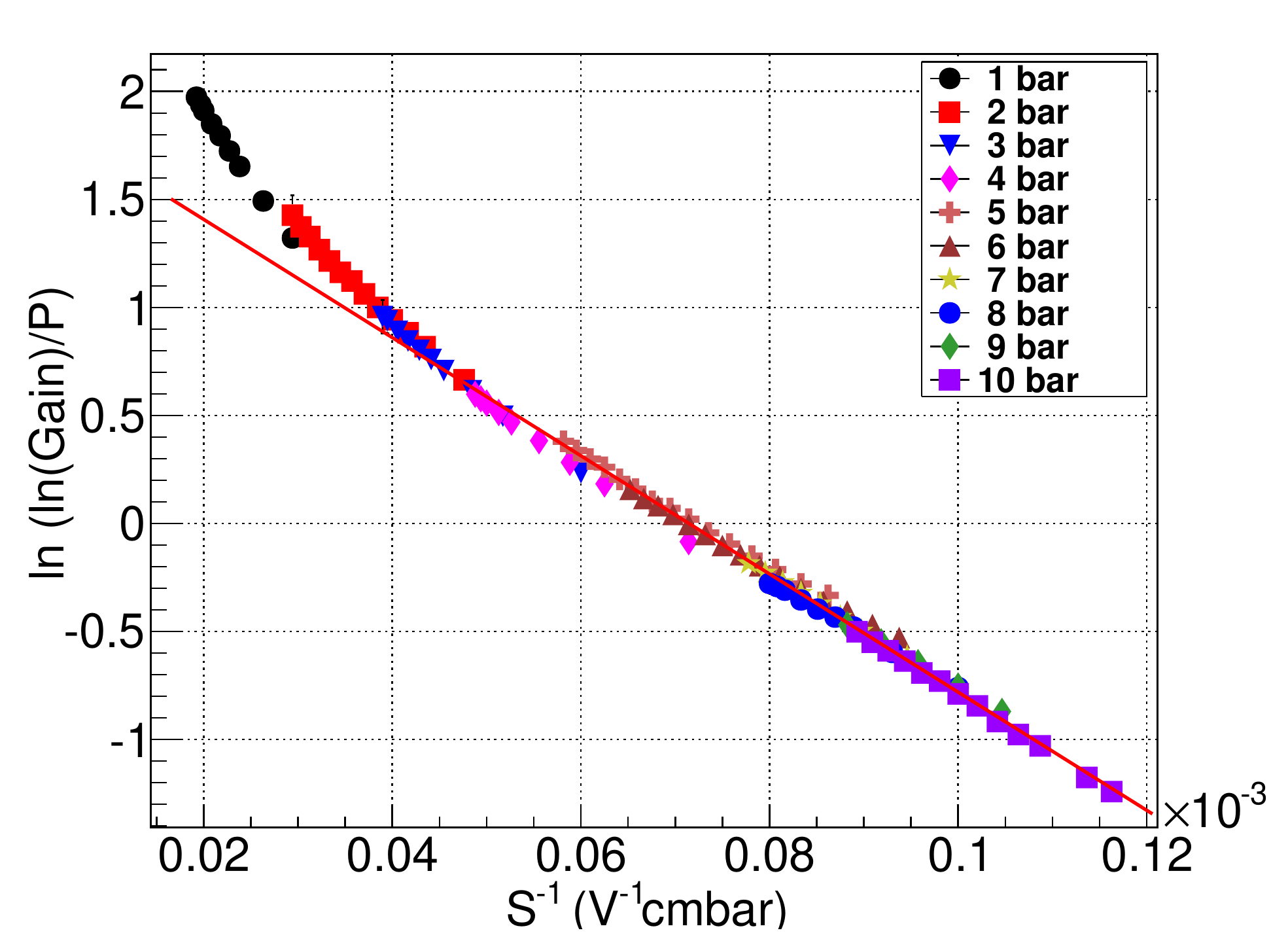}
\par\end{centering}
\caption{The natural logarithm of the reduced gas gain as a function of the inverse of the reduced amplification field. 
The red line represents a linear fit made for all data above $5$ bar, which emphasizes the deviations from the expected behaviour for data below this pressure. 
\label{fig:ParPlot}}
\end{figure}

\begin{figure}[h]
\begin{centering}
\includegraphics[scale=0.35]{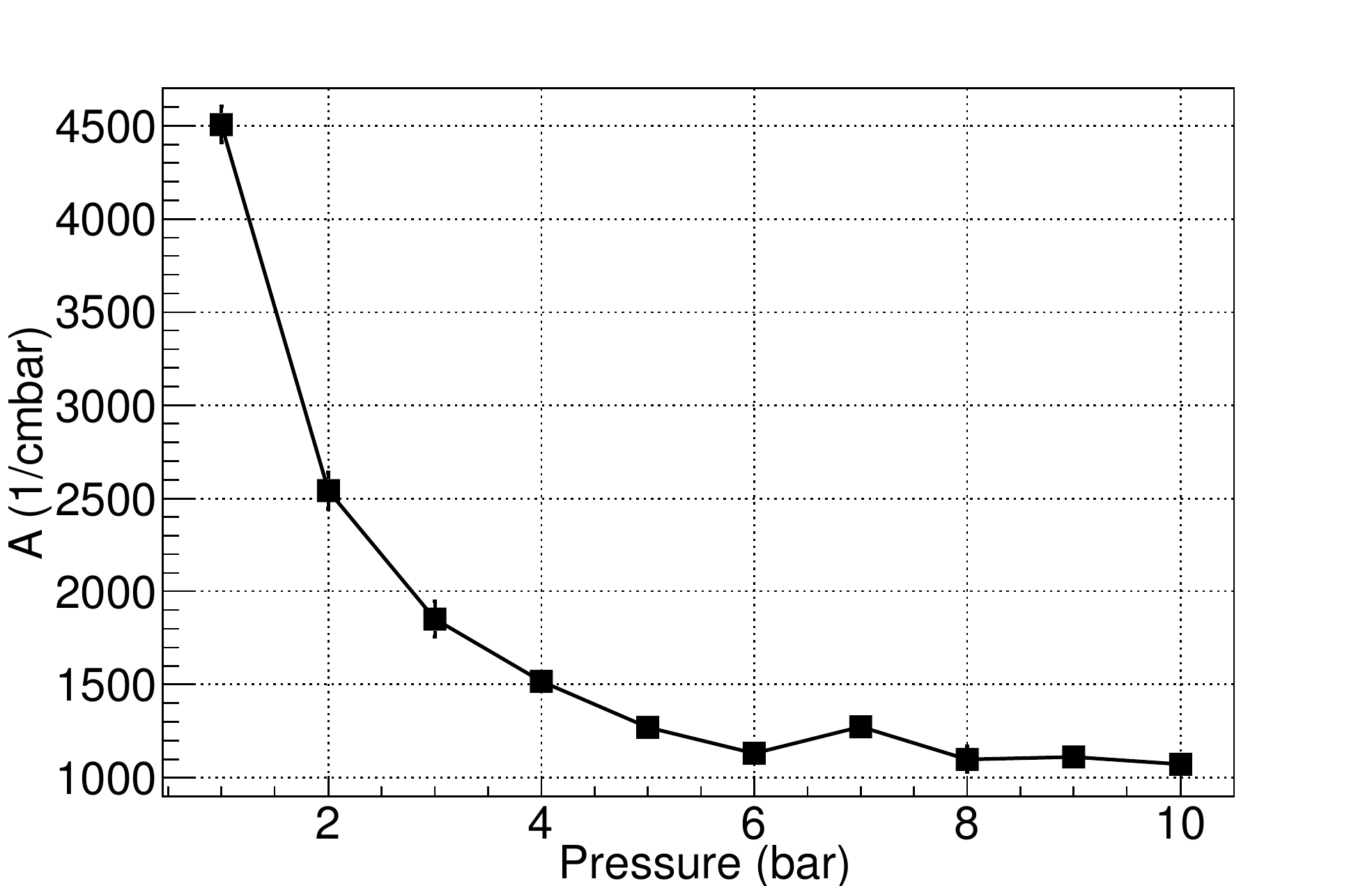}
\includegraphics[scale=0.35]{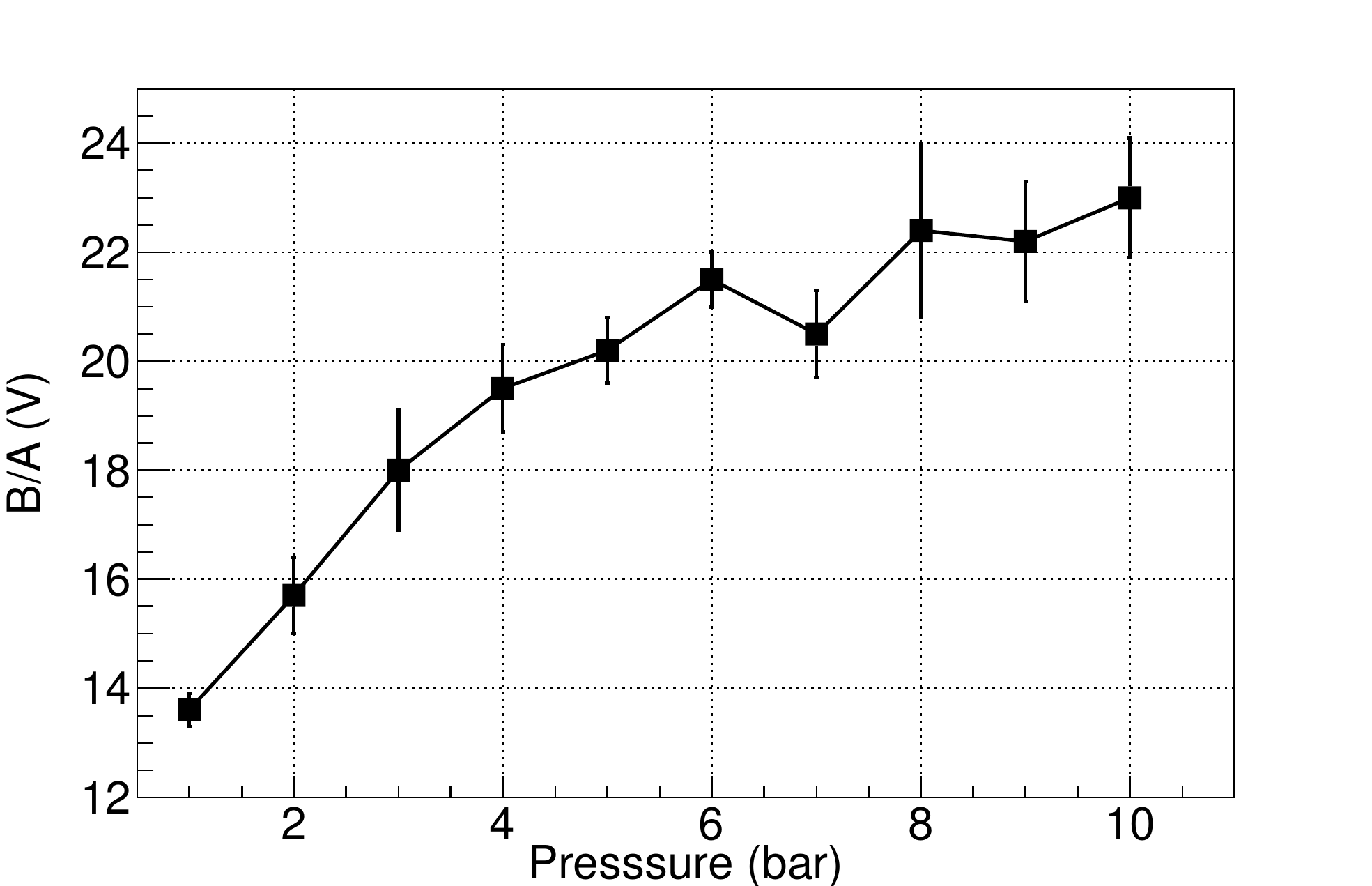}
\par\end{centering}
\caption{Parameter A (left) and the ratio of B to A (right) as a function of pressure. Parameters have been obtained from gas gain curves at each pressure, considering a semi-empirical formula of the first Townsend coefficient.   
\label{fig:ABParameter}}
\end{figure}

Figure \ref{fig:ParPlot} shows the value of ($\ln(\ln(G)/P)$) as a function of ($S^{-1}$), including data from all pressures measured.
Under the assumptions implicit in the model of equations \ref{eq:first} and \ref{eq:Main} we would expect all data points falling on the same straight line.
However, deviations from the model expectation are observed for pressures below $5$ bar, something that is more clearly observed in figure \ref{fig:ABParameter},
where the parameters $A$ (left) and the ratio $B/A$ (right) are plotted as a function of the pressure.
Both $A$ and $B$ were determined from linear fits of the curves showed in figure \ref{fig:ParPlot}.
As seen, $A$ drops with increasing pressure up to $5$ bar, remaining at a relatively constant value for higher pressures, 
while $B/A$ increases with pressure, reaching a stable value from $5$ bar on.
In contrast, measurements performed in cylindrical proportional detectors in the same pressure range, 
both in pure Xe and in Xe-methane mixtures \cite{Xeup10bar}, 
as well as measurements performed with Micromegas up to $2.5$ bar in Xe-methane \cite{XeMiTow}
have shown that the parameters $A$ and $B$ are independent of the pressure, 
and thus the model of equations \ref{eq:first} and \ref{eq:Main} represents a good description of those mixtures. 
We tentatively attribute the departure of our data from the expected model trends to the presence of Penning effect. 
However this fact needs further study, in particular using detailed Monte Carlo simulation of the avalanche microphysics, study that we expect to carry out in the near future.

\section{Conclusions}

We have operated microbulk Micromegas detectors in high pressure Xe+TMA Penning gas mixtures,
obtaining very good performance in terms of stability, gain and energy resolution. 
Systematic data on gain and energy resolution have been presented for a variety of TMA concentrations ranging from $0.3\%$ to $15\%$
and for gas pressures from $1$ to $10$ bar. We found that for TMA concentrations in the $1.5\%$-$2.5\%$ range, 
the Penning effect is maximum, requiring operating amplification fields at least $40\%$ 
lower than pure Xe for the same gain, or gains up to a factor $\sim$100 higher for the same amplification field. 
Maximum workable gains are superior by a factor at least $\times 3$ with respect to pure Xe, obtaining in particular gains above $400$ at $10$ bar.

For appropriate TMA fractions, energy resolutions down to $7.3\%$ ($9.6\%$) FWHM at $1$ ($10$) bar for the $22.1$ keV $^{109}$Cd peak could be achieved, 
an improvement of about a factor $2$ ($3$) with respect to values obtained by microbulks in pure Xe \cite{Coimbra}. 
This extrapolates into an energy resolution of $0.7\%$ ($0.9\%$) FWHM at the $Q_{\beta\beta}$ value of Xe for $1$ ($10$) bar, 
and therefore opens very good prospects for double-beta decay experiments.

In general, the performance of the detector is comparable to optimum values typically obtained with benchmark Micromegas mixtures (e.g. optimized Ar-isobutane mixtures). 
This result proves that Xe+TMA may be a mixture of choice for Micromegas in applications envisaging the use of Xe as conversion gas, especially at high pressures.

\acknowledgments
We are grateful to our colleagues of the groups of the University of
Zaragoza, CEA/Saclay and our colleagues from the NEXT and RD-51
collaborations for helpful discussions and encouragement. We thank R.
de Oliveira and his team at CERN for the manufacturing of the
microbulk readouts. We acknowledge support from the European
Commission under the European Research Council T-REX Starting Grant
ref. ERC-2009-StG-240054 of the IDEAS program of the 7th EU Framework
Program. We also acknowledge support from the Spanish Ministry of Economics and Competitiveness (MINECO), 
under contracts ref. FPA2008-03456 and FPA2011-24058, as
well as under the CUP project ref. CSD2008-00037 and the CPAN project
ref. CSD2007- 00042 from the Consolider-Ingenio 2010 program of the
MICINN. Part of these grants are funded by the European Regional
Development Fund (ERDF/FEDER). 
F.I. acknowledges the support from the Eurotalents program and 
D.C.H. of the Univ. Zaragoza, under the
program PIF-UZ-2009-CIE-03.

\end{document}